\documentclass[a4paper,10pt]{article}
\usepackage[utf8]{inputenc}
\usepackage[T1]{fontenc}
\usepackage{geometry}
\usepackage{microtype}
\usepackage[scaled=0.92]{helvet}
\usepackage{palatino,mathpazo}
\usepackage[scaled=1.02]{inconsolata}
\usepackage{amsmath,amssymb,array}
\usepackage{booktabs,longtable}
\usepackage{graphicx}
\usepackage{float}
\usepackage{placeins}
\usepackage[medium]{titlesec}
\titleformat{\section}{\normalfont\large\bfseries}{\arabic{section}}{1em}{}
\titleformat{\subsection}{\normalfont\normalsize\bfseries}{\arabic{section}.\arabic{subsection}}{0.5em}{}
\usepackage[round,authoryear]{natbib}
\usepackage[hyphens]{url}
\usepackage[pagebackref]{hyperref}

\usepackage{color}
\definecolor{link}{rgb}{0.45,0.51,0.67}
\hypersetup{colorlinks,citecolor=link,filecolor=link,linkcolor=link,urlcolor=link}
\usepackage{fancyvrb}
\usepackage[font=small,labelfont=bf]{caption}
\usepackage{setspace}
\newcommand{\pkg}[1]{\textbf{#1}}
\newcommand{\CRANpkg}[1]{\href{https://CRAN.R-project.org/package=#1}{\pkg{#1}}}

\renewenvironment{abstract}{\noindent\textbf{Abstract}~}{\par}
\title{\pkg{marp}: Model-Averaged Inference for Parametric Renewal Processes in R}
\author{Veronica W. Tsou$^{1,2}$, Vanessa Huang$^{1}$, and Jie Kang$^{1,2,3}$\\[0.5em]\normalfont\small $^{1}$School of Mathematics and Statistics, The University of Sydney, Camperdown 2050, Australia\\$^{2}$Charles Perkins Centre, The University of Sydney, Camperdown 2050, Australia\\$^{3}$Sydney Precision Data Science Centre, The University of Sydney, Camperdown 2050, Australia\\[0.3em]Corresponding author: Jie Kang (\href{mailto:jie.kang@sydney.edu.au}{jie.kang@sydney.edu.au})}
\date{}

\makeatletter
\renewcommand{\maketitle}{%
  \noindent{\huge\bfseries\raggedright \@title\par}
  \vspace{0.5em}
  \noindent\textit{\@author}\par
  \vspace{0.7em}
  \noindent\@date\par
  \vspace{1em}
}
\makeatother
\begin{document}
\maketitle
\begin{abstract}
Renewal processes model inter-event times between recurrent events using a parametric distribution, but selecting a single model can ignore model uncertainty and affect inference, particularly in small samples. This paper introduces \CRANpkg{marp}, an R package for model-averaged inference in parametric renewal processes. The package fits six commonly used distributions, compares them using Akaike and Bayesian information criteria, and reports selected-model and model-averaged estimates of the mean inter-event time, event probability and hazard function, with percentile and Studentised bootstrap confidence intervals. A simulation study under three data-generating models showed that model averaging did not improve point estimation but often improved interval coverage relative to the AIC-selected model. Applications to seizure-recurrence and engineering failure-time data illustrate the proposed methods. The \CRANpkg{marp} package provides a reproducible framework for inference that accounts for model uncertainty in parametric renewal processes.
\end{abstract}

\section{Introduction}\label{introduction}

Renewal processes are widely used to model recurrent events through the distribution of inter-event times \citep{smith1958renewal, ross1996stochastic, cook2007statistical}. Common choices include the exponential, gamma, Weibull, log-normal, log-logistic and Brownian passage-time (BPT) distributions, which accommodate hazard functions ranging from the constant \citep{lawless2003statistical} to flexible monotone \citep{lawless2003statistical, klein2003survival} and non-monotone forms \citep{lawless2003statistical, bennett1983log}. Candidate models are typically fitted by maximum likelihood and selected using the Akaike Information Criterion (AIC) \citep{akaike1973information} or Bayesian Information Criterion (BIC) \citep{schwarz1978estimating}.

When the data do not clearly favour any particular model, inference based on a single selected model ignores model uncertainty and can affect point and interval estimation, particularly in small samples \citep{chatfield1995model, draper1995assessment, buckland1997model}. Frequentist model averaging addresses this by combining model-specific estimates using weights derived from the relative support for each candidate model \citep{hjort2003frequentist, claeskens2012model}.

Model averaging has been studied extensively in regression, prediction and ecological inference \citep{burnham2002model, hansen2007least, claeskens2012model}, but its application to renewal processes remains limited. In renewal-process models, model averaging combines estimates of common estimands, such as the mean inter-event time, event probabilities and hazard function, across multiple candidate waiting-time distributions to account for model uncertainty. This motivates the development of a unified framework for model-averaged inference in renewal processes.

This paper introduces \CRANpkg{marp} (\emph{Model Averaging for Renewal Processes}), an \texttt{R} package for model-averaged inference in renewal processes. The package currently implements exponential, gamma, Weibull, log-normal, log-logistic and Brownian passage-time (BPT) models. For each fitted model, it estimates common estimands, including the mean inter-event time, event probabilities at user-specified times and the hazard function, and reports both selected-model and information-criterion-weighted model-averaged estimates. \CRANpkg{marp} also provides percentile and Studentised bootstrap confidence intervals. Bootstrap model-selection frequencies are also returned as descriptive measures of selection stability.

The contribution of \CRANpkg{marp} is a unified framework for model-averaged inference across commonly used parametric renewal-process models. The simulation study evaluates the finite-sample estimation and confidence-interval performance under correctly specified and misspecified models, and two applications illustrate the effect of model uncertainty on substantive inference.

\section{Statistical framework}\label{sec:framework}

\subsection{Renewal-process setting}\label{sec:setting}

Let \(\{N(t): t \ge 0\}\) be a renewal process with event times
\(0 < T_1 < T_2 < \cdots\) and inter-event times
\(X_i = T_i - T_{i-1}\), where \(T_0 = 0\).
A renewal process is characterised by the assumption that \(\{X_i\}_{i \ge 1}\) are independent and identically distributed
non-negative random variables. Let \(F\), \(f\) and \(S=1-F\) denote the distribution function, density and survival function of an inter-event time (X), respectively. The hazard function is
\[
h(t)=\frac{f(t)}{S(t)}, \qquad t>0.
\]
In this setting, the waiting-time distribution \(F\) is the primary modelling object \citep{smith1958renewal}. A parametric renewal-process model assumes that \(F\) belongs to a family \(F(\cdot;\theta)\), where \(\theta\in\Theta\) is a finite-dimensional parameter vector. Estimation is based on an observed numeric vector of inter-event times,
\[
\mathbf{X}=(x_1,\ldots,x_n).
\]

The framework considered here assumes complete, uncensored inter-event times that are independent and identically distributed. Censoring covariates, subject-level heterogeneity and dependence between successive waiting times are outside the scope of the current implementation.

\subsection{Candidate models}\label{sec:candidate-dist}

The candidate model set in \CRANpkg{marp} consists of six parametric waiting-time distributions: exponential, gamma, Weibull, log-normal, log-logistic and Brownian passage time (BPT). These distributions provide a range of possible inter-event-time behaviours and hazard shapes, from the constant hazard implied by the exponential model to more flexible monotone or non-monotone hazards under the remaining candidate models. The exponential waiting-time model corresponds to a homogeneous Poisson process. However, the distribution fitted to the observed inter-event times is exponential, not Poisson, because a Poisson distribution models event counts rather than waiting times.

Let \(M_1,\ldots,M_K\) denote the candidate waiting-time models. Under model \(M_k\), the inter-event time distribution is written as
\[
F_k(t;\boldsymbol{\theta}_k),
\]
where \(\boldsymbol{\theta}_k\in\Theta_k\) is the model-specific parameter vector. The corresponding density, survival function and hazard function are denoted by
\[
f_k(t;\boldsymbol{\theta}_k), \qquad
S_k(t;\boldsymbol{\theta}_k)=1-F_k(t;\boldsymbol{\theta}_k),
\]
and
\[
h_k(t;\boldsymbol{\theta}_k)
=
\frac{f_k(t;\boldsymbol{\theta}_k)}{S_k(t;\boldsymbol{\theta}_k)}.
\]
The model-specific parameter vector is estimated by maximum likelihood. If \(\ell_k(\boldsymbol{\theta}_k)\) denotes the log-likelihood under model \(M_k\), then the maximum likelihood estimator is
\[
\widehat{\boldsymbol{\theta}}_k
=
\arg\max_{\boldsymbol{\theta}_k\in\Theta_k}
\ell_k(\boldsymbol{\theta}_k).
\]
The BPT model is included as a flexible right-skewed waiting-time model and is closely related to the inverse-Gaussian first-passage-time distribution. Under a mean--shape inverse-Gaussian parameterisation, the density
\[
f(t;\mu,\lambda)
=
\left(\frac{\lambda}{2\pi t^3}\right)^{1/2}
\exp\left\{
-\frac{\lambda(t-\mu)^2}{2\mu^2t}
\right\},
\qquad
t>0,\quad \mu>0,\quad \lambda>0.
\]
Under this parameterisation, \(\mu\) is the mean parameter and \(\lambda\) is a positive shape parameter, with
\[
\mathrm{E}(X)=\mu,\qquad \mathrm{var}(X)=\frac{\mu^3}{\lambda}.
\]
Because inverse-Gaussian and Brownian first-passage-time models are sometimes written under different but equivalent parameterisations, the parameterisation is stated explicitly \citep{chhikara1977inverse, folks1978inverse}. In \CRANpkg{marp}, the BPT parameters are constrained to be positive, and model-specific starting values are used to improve numerical stability during maximum likelihood estimation.

\subsection{Estimands}\label{sec:functional}

The analysis focuses on estimands that are common to all candidate waiting-time models. The main quantities are the mean inter-event time,
\[
\mu = \mathrm{E}(X),
\]
the event probability, or distribution function, at a threshold \(y\),
\[
F(y)=Pr(X\leq y),
\]
and the hazard function at selected time points \(t\),
\[
h(t)=\frac{f(t)}{S(t)}.
\]

For model \(M_k\), the corresponding plug-in estimators are
\[
\widehat{\mu}_k=\mu_k(\widehat{\boldsymbol{\theta}}_k),
\]
\[
\widehat{F}_k(y)=F_k(y;\widehat{\boldsymbol{\theta}}_k),
\]
and
\[
\widehat{h}_k(t)=h_k(t;\widehat{\boldsymbol{\theta}}_k).
\]
In particular, the transformed event-probability scale is
\[
\operatorname{logit}\{\widehat{F}_k(y)\}
=
\log\left\{
\frac{\widehat{F}_k(y)}
{1-\widehat{F}_k(y)}
\right\},
\]
and the transformed hazard scale is
\[
\log\{\widehat{h}_k(t)\}.
\]
In the \CRANpkg{marp} documentation, \texttt{pr\_hat} is described as an estimated logit probability and \texttt{haz\_hat} as an estimated log hazard rate. Therefore, this paper distinguishes between the natural event probability \(F(y)\), the logit event probability \(\operatorname{logit}\{F(y)\}\), the hazard \(h(t)\), and the log-hazard \(\log\{h(t)\}\). This distinction is important because averaging hazards on the natural scale is not equivalent to averaging log-hazards and transforming back. Similarly, averaging probabilities on the natural scale is not equivalent to averaging logit probabilities and transforming back.

\subsection{Selected-model and model-averaged point estimation}\label{sec:ic-avg}

Let \(M_1,\ldots,M_K\) denote the candidate waiting-time models. For model \(M_k\), let \(\widehat{\boldsymbol{\theta}}_k\) be the maximum likelihood estimator, let \(\ell_k(\widehat{\boldsymbol{\theta}}_k)\) be the maximised log-likelihood and let \(d_k\) be the number of estimated parameters. The Akaike information criterion (AIC) is calculated as
\[
\mathrm{AIC}_k
=
-2\ell_k(\widehat{\boldsymbol{\theta}}_k)+2d_k.
\]
The minimum-AIC model is
\[
\widehat{k}_{\min}
=
\arg\min_{1\leq k\leq K}\mathrm{AIC}_k.
\]
For a target functional \(\psi\), let \(\widehat{\psi}_k\) denote the plug-in estimator under model \(M_k\). The selected-model estimator is
\[
\widehat{\psi}_{\mathrm{sel}}
=
\widehat{\psi}_{\widehat{k}_{\min}}.
\]
AIC weights are computed as
\[
w_k
=
\frac{\exp(-\Delta_k/2)}{\sum_{j=1}^{K}\exp(-\Delta_j/2)},
\qquad
\Delta_k
=
\mathrm{AIC}_k-\min_j \mathrm{AIC}_j.
\]
These weights measure relative empirical support within the specified candidate model set. They are used for model-averaging weights but should not be interpreted as posterior probabilities that the models are true. The AIC-weighted model-averaged estimator of \(\psi\) is
\[
\widehat{\psi}_{\mathrm{MA}}
=
\sum_{k=1}^{K}w_k\widehat{\psi}_k.
\]
Examples include the model-averaged mean inter-event time,
\[
\widehat{\mu}_{\mathrm{MA}}
=
\sum_{k=1}^{K}w_k\widehat{\mu}_k,
\]
the event probability,
\[
\widehat{F}_{\mathrm{MA}}(y)
=
\sum_{k=1}^{K}w_k\widehat{F}_k(y),
\]
the hazard function,
\[
\widehat{h}_{\mathrm{MA}}(t)
=
\sum_{k=1}^{K}w_k\widehat{h}_k(t).
\]
Averaging on a transformed scale defines a different estimand. For example, averaging log-hazards gives
\[
\widehat{\eta}_{h,\mathrm{MA}}(t)
=
\sum_{k=1}^{K}w_k\log\{\widehat{h}_k(t)\},
\]
which is not equivalent to \(\log\{\widehat{h}_{\mathrm{MA}}(t)\}\). Therefore, the software implementation and simulation study distinguish between natural- and transformed-scale summaries.

\subsection{Bootstrap confidence intervals}\label{sec:bootstrap}

Bootstrap inference is used to quantify uncertainty in both selected-model and model-averaged estimators. Bootstrap resampling is applied to the entire estimation procedure, with candidate models refitted, information criteria recalculated, the selected model reidentified and model-averaged estimates recomputed within each resample. Confidence intervals for \(\widehat{\psi}_{\mathrm{MA}}\) are directly obtained from its bootstrap distribution, thereby incorporating both sampling variability and model-selection uncertainty.

Let \(\widehat{\psi}\) denote an estimator of a target quantity \(\psi\). For each outer bootstrap resample \(b=1,\ldots,B\), a bootstrap sample
\(\mathbf{x}^{*(b)}=(x_1^{*(b)},\ldots,x_n^{*(b)})\)
is drawn with replacement from the observed inter-event times \(x=(x_1,\ldots,x_n)\). Candidate models are refitted to \(x^{*(b)}\), AIC values and AIC weights are recalculated, the minimum-AIC model is reselected, and selected-model and model-averaged estimates are recomputed. Repeating this procedure across bootstrap resamples yields a bootstrap distribution that incorporates both sampling variability and model-selection uncertainty.

Percentile bootstrap intervals are obtained from empirical quantiles of the bootstrap estimates. Studentised bootstrap intervals additionally require an estimate of the standard error within each outer bootstrap resample. For an estimator \(\widehat{\psi}\), the Studentised bootstrap statistic can be written as
\[
Z_b^*
=
\frac{\widehat{\psi}_b^*-\widehat{\psi}}
{\widehat{\mathrm{se}}_b^*}.
\]
Here, \(\widehat{\psi}_b^*\) is the estimate from the \(b\)-th outer bootstrap sample, and \(\widehat{\mathrm{se}}_b^*\) is its standard-error estimate obtained from inner bootstrap resamples. In the double-bootstrap procedure, \(B_B\) inner bootstrap samples are drawn within each outer resample. The full fitting and model-averaging workflow is repeated for these inner resamples, and the empirical standard deviation of the resulting estimates is used as \(\widehat{\mathrm{se}}_b^*\).

\section{Software implementation}\label{sec:implementation}

The \CRANpkg{marp} package implements the statistical framework described in \hyperref[sec:framework]{Section 2} in \texttt{R} \citep{rcoreteam2026}. It provides a unified framework for fitting parametric renewal-process models, performing information-criterion model averaging, and constructing bootstrap confidence intervals.

\subsection{Package architecture}\label{sec:architecture}

The implementation has two levels. First, model-specific fitting functions estimate individual candidate waiting-time distributions. Second, wrapper functions combine these fitted models to model averaging, selected-model inference and bootstrap inference.

The model-specific functions are \texttt{poisson\_rp()}, \texttt{gamma\_rp()}, \texttt{weibull\_rp()}, \texttt{lognorm\_rp()}, \texttt{loglogis\_rp()} and \texttt{bpt\_rp()}. The function \texttt{poisson\_rp()} fits an exponential waiting-time model and is labelled ``Poisson'' because exponential inter-event times correspond to a homogeneous Poisson process.

Each model-specific function takes a numeric vector of observed inter-event times and returns a common set of outputs, including parameter estimates, likelihood-based fit summaries, AIC and BIC values, and estimated mean inter-event time, event-probability summaries and hazard or log-hazard summaries at user-specified time points. Returning these quantities in a common format enables model averaging and selected-model inference to be applied uniformly across all candidate models.

The primary wrapper functions are \texttt{marp()} and \texttt{confint()}. The function \texttt{marp()} fits the candidate models and returns information criteria, model weights and AIC-weighted model-averaged estimates. The function \texttt{confint()} extends this framework by constructing percentile and Studentised bootstrap confidence intervals through repeated model fitting, model selection and model averaging within bootstrap resamples.

\subsection{User interface and returned objects}\label{sec:interface}

The primary point-estimation function is \texttt{marp(data,\ t,\ m,\ y,\ which.model)}.
Here, \texttt{data} is a numeric vector of observed inter-event times, \texttt{t} specifies the evaluation time(s) for hazard summaries, \texttt{m} controls optimisation settings, \texttt{y} is the threshold at which the event probability \(Pr(X\leq y)\) is estimated, and \texttt{which.model} identifies a known reference for simulation studies. In practice, inference is based on the fitted candidate models, the minimum-AIC model and the corresponding model-averaged estimates.

Bootstrap confidence intervals are obtained using \texttt{confint(object,\ data,\ B,\ BB,\ level)}.
This function implements percentile and Studentised bootstrap intervals. The arguments \texttt{B} and \texttt{BB} control the number of outer and inner bootstrap samples, respectively, while \texttt{level} determines the nominal confidence level.

The returned objects are designed to support both direct reporting and downstream analyses. Outputs include model-specific parameter estimates, information criteria and model weights, selected-model and model-averaged estimates, bootstrap confidence intervals, and bootstrap model-selection frequencies.

Natural- and transformed-scale quantities are returned separately. For example, event probabilities are distinguished from logit event probabilities, and hazards from log-hazards, because model averaging on a transformed scale is generally not equivalent to transforming a natural-scale model-averaged estimate.

Further details of the model-specific fitting functions, returned objects and bootstrap outputs are provided in the Supplementary Material (Section 1).

\subsection{Typical workflow}\label{sec:workflow}

A typical \CRANpkg{marp} analysis starts with a numeric vector of complete inter-event times. The user chooses the time points at which to evaluate the hazard function, a threshold for the event probability, and a value of \texttt{m} controlling repeated optimisation starts for numerically challenging candidate models. The main fitting call then fits all six candidate waiting-time distributions, computes information-criterion weights and returns both selected-model and model-averaged summaries.

\begin{verbatim}
library(marp)

# Complete inter-event times from one renewal process.
dat <- c(
  8.2, 12.5, 18.4, 21.7, 27.3, 33.6, 41.2, 45.8,
  53.9, 64.1, 78.5, 91.4, 112.6, 136.8, 171.3
)

# Analysis settings.
t_eval <- as.numeric(
  quantile(dat, probs = c(0.25, 0.50, 0.75))
)
y_eval <- mean(dat)
m <- 50

fit <- marp::marp(data = dat, t = t_eval, m = m, y = y_eval)
fit
\end{verbatim}

\begin{verbatim}
#> Model-averaged renewal-process fit
#> Observations: 15 
#> 
#> Model comparison:
#>                                        Model      AIC AIC weight Status
#>  Poisson renewal (exponential waiting times) 155.4015  0.1844565     ok
#>                                        Gamma 155.3311  0.1910720     ok
#>                                 Log-logistic 156.5081  0.1060727     ok
#>                                      Weibull 155.5266  0.1732740     ok
#>                                   Log-normal 155.5637  0.1700932     ok
#>                        Brownian passage time 155.5064  0.1750315     ok
#> 
#> Best model: Gamma 
#> Best-model mean: 61.15333 
#> Model-averaged mean: 62.83834 
#> Model-averaged logit event probability: 0.5305688 
#> Model-averaged log-hazard evaluations: 3 stored
\end{verbatim}

The fitted object can be used directly for reporting. For example, \texttt{model\_best} identifies the minimum-AIC model, while \texttt{mu\_best}, \texttt{pr\_best} and \texttt{haz\_best} contain the selected-model estimates. The corresponding AIC-weighted model-averaged summaries are \texttt{mu\_aic}, \texttt{pr\_aic} and \texttt{haz\_aic}.

\begin{verbatim}
fit_summary <- summary(fit)
fit_summary
\end{verbatim}

\begin{verbatim}
#> Model-averaged renewal-process fit summary
#> Observations: 15 
#> 
#> Model comparison:
#>                                        model        par1        par2    logLik
#>  Poisson renewal (exponential waiting times)  0.01635234          NA -76.70077
#>                                        Gamma  1.67622993  0.02741028 -75.66553
#>                                 Log-logistic  1.96661975 45.39801577 -76.25405
#>                                      Weibull 66.88512363  1.34325647 -75.76330
#>                                   Log-normal  3.78635719  0.85798080 -75.78183
#>                        Brownian passage time 61.15333213  1.00846773 -75.75321
#>       AIC      BIC AIC_weight status
#>  155.4015 156.1096  0.1844565     ok
#>  155.3311 156.7472  0.1910720     ok
#>  156.5081 157.9242  0.1060727     ok
#>  155.5266 156.9427  0.1732740     ok
#>  155.5637 156.9798  0.1700932     ok
#>  155.5064 156.9225  0.1750315     ok
#> 
#> Selected model: Gamma 
#> Selected-model mean: 61.15333 
#> Selected-model logit event probability: 0.4162729 
#> 
#> Model-averaged estimates:
#> Mean: 62.83834 
#> Logit event probability at y = 61.15333 : 0.5305688 
#> Log-hazard evaluations: 3 stored
#> 
#> Reference model: Poisson renewal (exponential waiting times)
\end{verbatim}

\begin{verbatim}
c(
  selected_mean = fit$mu_best,
  model_averaged_mean = fit$mu_aic
)
\end{verbatim}

\begin{verbatim}
#>       selected_mean model_averaged_mean 
#>            61.15333            62.83834
\end{verbatim}

\begin{verbatim}
data.frame(
  time = t_eval,
  selected_log_hazard = fit$haz_best,
  model_averaged_log_hazard = fit$haz_aic
)
\end{verbatim}

\begin{verbatim}
#>    time selected_log_hazard model_averaged_log_hazard
#> 1 24.50           -4.181544                 -4.047543
#> 2 45.80           -3.975254                 -3.956079
#> 3 84.95           -3.830028                 -3.967061
\end{verbatim}

Uncertainty can then be quantified by applying \texttt{confint()} to the same analysis settings. The small bootstrap settings below keep the example lightweight; final analyses should use larger values of \texttt{B} and \texttt{BB} because Studentised intervals involve a nested bootstrap.

\begin{verbatim}
set.seed(2026)

ci <- confint(
  fit,
  data = dat,
  B = 49,
  BB = 19,
  level = 0.95
)

c(
  lower = ci$student_CI$mu_lower_ma,
  estimate = fit$mu_aic,
  upper = ci$student_CI$mu_upper_ma
)
\end{verbatim}

\begin{verbatim}
#>     lower  estimate     upper 
#>  40.69106  62.83834 133.64858
\end{verbatim}

\begin{verbatim}
hazard_summary <- data.frame(
  time = t_eval,
  log_hazard = fit$haz_aic,
  lower = ci$student_CI$haz_lower_ma,
  upper = ci$student_CI$haz_upper_ma
)

plot(
  hazard_summary$time,
  hazard_summary$log_hazard,
  type = "l",
  xlab = "Time",
  ylab = "Model-averaged log-hazard"
)
lines(hazard_summary$time, hazard_summary$lower, lty = 2)
lines(hazard_summary$time, hazard_summary$upper, lty = 2)
\end{verbatim}

\begin{center}\includegraphics[width=1\linewidth]{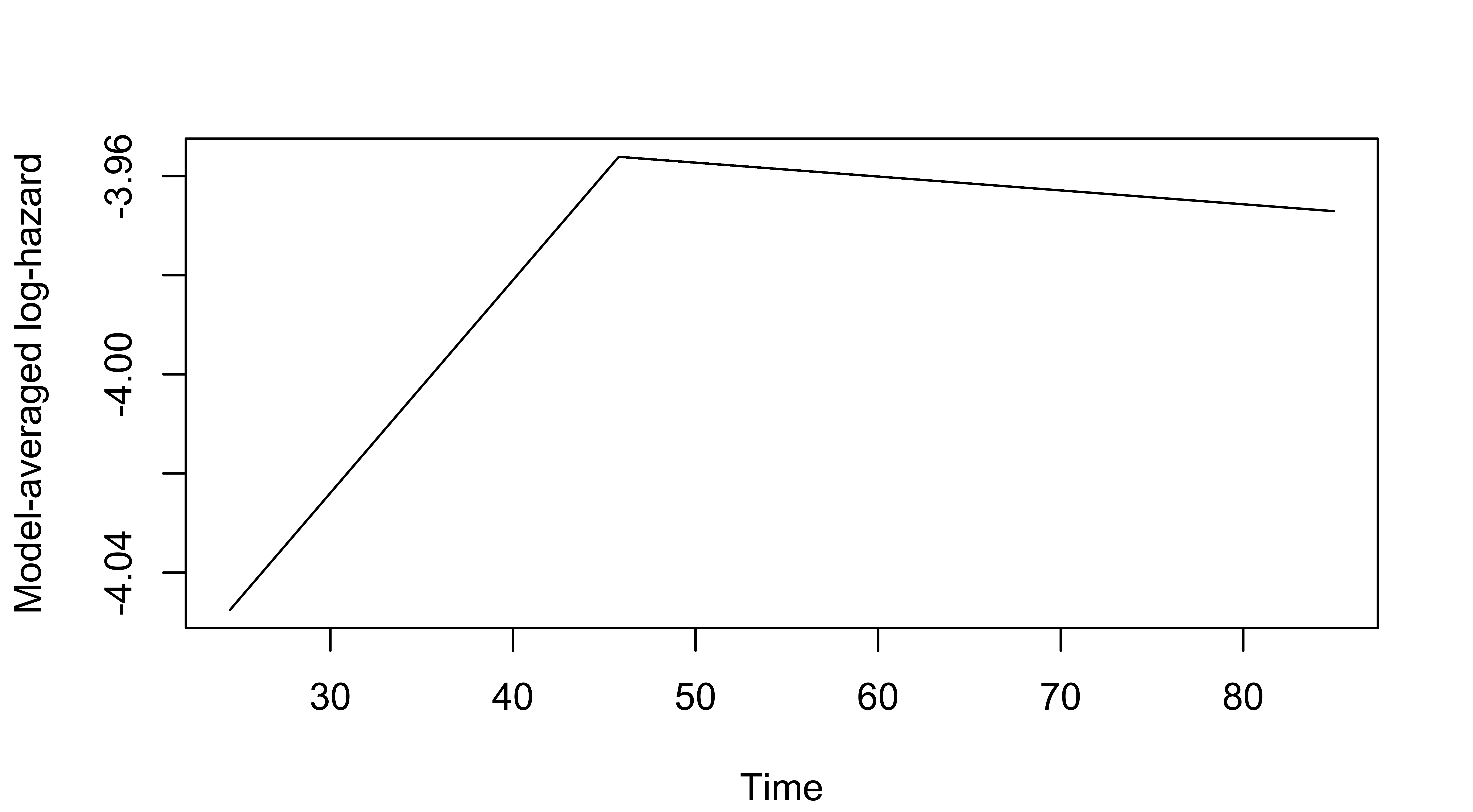} \end{center}

\subsection{Optimisation, starting values and failed fits}\label{sec:optimisation}

The model-specific functions use maximum likelihood estimation to fit each candidate waiting-time distribution. Because the candidate distributions have different parameter constraints and likelihood surfaces, numerical optimisation is performed on transformed scales where appropriate.

Initial values are generated using model-specific rules based on empirical summaries of the observed inter-event times, with multiple random restarts for numerically challenging models, particularly the BPT distribution. These strategies improve optimisation robustness for small samples and highly skewed waiting-time distributions.

The \texttt{bpt\_rp()} function includes an additional plausibility check because optimisation of the BPT likelihood can be unstable in small or highly skewed samples. Candidate BPT fits with estimated mean inter-event times exceeding \(10\max_i x_i\), where \(x_i\) denotes the observed inter-event times, are rejected and optimisation is restarted. The number of attempts is capped at \texttt{max\_attempts\ \textless{}-\ m\ *\ 20}. If no plausible solution is obtained, the function returns a controlled failed fit with missing parameter estimates and infinite AIC and BIC values, allowing the remaining candidate models to be analysed.

Failed candidate fits are assigned zero AIC weight and therefore do not contribute to model-averaged estimates. If no finite-AIC candidate model is available, the corresponding model fit is treated as failed. When a target estimand or confidence-interval bound cannot be computed reliably, the corresponding estimate or bound is returned as missing, and the failed confidence interval is recorded in the simulation summaries.

\subsection{Comparison with existing software}\label{sec:comparison}

Several \texttt{R} packages provide parts of the workflow needed for parametric time-to-event analysis. Distribution-fitting packages such as \CRANpkg{fitdistrplus} \citep{delignettemuller2015fitdistrplus} can fit many univariate distributions and assess goodness of fit, while survival-analysis packages such as \CRANpkg{survival} \citep{therneau2000survival, therneau2026survival} and \CRANpkg{flexsurv} \citep{jackson2016flexsurv} provide broad support for censored survival models, regression and prediction. General model-averaging packages such as \CRANpkg{FMAdist} \citep{nelson2022fmadist} and \CRANpkg{MuMIn} \citep{barton2026mumin} address model averaging, but they are not organised around renewal-process estimands.

The contribution of \CRANpkg{marp} is different. It provides a single workflow for complete inter-event-time data in which the same six parametric waiting-time models are fitted, compared and averaged for renewal-process quantities. In particular, \texttt{marp()} returns selected-model and AIC-weighted estimates of the mean inter-event time, event probability and log-hazard at user-specified time points, while \texttt{confint()} repeats the full fitting, model-selection and model-averaging procedure inside bootstrap resampling. Without this package, users would need to fit candidate distributions separately, calculate information criteria and weights by hand, extract comparable estimands on consistent scales, combine model-specific estimates, and reimplement these steps inside a nested bootstrap.

The closest existing package is \CRANpkg{FMAdist}, which, like \CRANpkg{marp}, averages over candidate distributions fitted to univariate data rather than over regression models. It differs in how the weights are obtained: \CRANpkg{FMAdist} uses the jackknife model averaging (JMA) method, in which the data are divided into \(J\) groups and the weights are chosen by cross-validation \citep{nelson2021input}. This is unreliable for renewal-process inference because time-to-event datasets typically have few observations, so the cross-validated weights are themselves estimated with substantial variability. The \CRANpkg{FMAdist} package also returns an averaged input distribution rather than the renewal-process estimands of interest, and provides no interval estimation of the kind supplied by \texttt{confint()}.

The \CRANpkg{MuMIn} package offers model selection and averaging through a range of information criteria, including the AIC used by \CRANpkg{marp}, but it operates on fitted model objects supporting \texttt{logLik()}, \texttt{coef()} and \texttt{vcov()} methods, and builds its candidate set from predictor subsets of a global model, so a set of competing waiting-time distributions is not a candidate set it can consume. Its unconditional standard errors are analytic Wald quantities rather than the Studentised bootstrap intervals returned by \texttt{confint()}. The \CRANpkg{AICcmodavg} package \citep{mazerolle2025aiccmodavg} likewise provides general information-criterion averaging, and although the two packages differ in their input requirements, in how the candidate model set is constructed and in the averaging methods offered, neither provides the renewal-model fitting nor the bootstrap interval estimation that \CRANpkg{marp} supplies.

Finally, \CRANpkg{fitdistrplus} fits both non-censored and censored univariate data, whereas \CRANpkg{marp} can only be used for complete (non-censored) inter-event times. Unlike \CRANpkg{marp}, however, \CRANpkg{fitdistrplus} has no model-averaging functionality, and its bootstrap confidence intervals apply to the parameters and quantiles of a single fitted distribution rather than to model-averaged estimands.

Table \ref{tab:model-comparison-pdf} summarises the practical differences. The table is not intended to claim that these packages solve identical problems; rather, it identifies where existing software provides adjacent functionality and where \CRANpkg{marp} supplies an integrated renewal-process workflow.

\begin{table}[!htbp]
\centering
\caption{\label{tab:model-comparison-pdf}Comparison of \CRANpkg{marp} with selected adjacent R packages: fitdistrplus, survival, flexsurv, FMAdist and MuMIn. The comparison focuses on the workflow needed for model-averaged inference from complete inter-event-time data. AICcmodavg is discussed in the text but omitted from the table.}
\centering
\fontsize{7}{9}\selectfont
\begin{tabular}[t]{>{\raggedright\arraybackslash}p{2.4cm}>{\raggedright\arraybackslash}p{1.7cm}>{\raggedright\arraybackslash}p{1.7cm}>{\raggedright\arraybackslash}p{1.95cm}>{\raggedright\arraybackslash}p{1.7cm}>{\raggedright\arraybackslash}p{1.7cm}}
\toprule
 & marp & fitdistrplus & survival/flexsurv & FMAdist & MuMIn\\
\midrule
Primary data setting & Univariate inter-event times & Complete or censored data & Survival data & Univariate data & Fitted model objects\\
Candidate model set & Six renewal models, inc. BPT & User-specified families & Survival models & 11 common distributions, inc. empirical & Predictor subsets of a global model\\
Model-comparison support & AIC and BIC & AIC, BIC and goodness-of-fit & AIC and BIC from fitted objects & Cross-validation & AICc and variants\\
Model averaging & AIC and bootstrap weights & Not provided & Not provided & JMA weights & Bootstrap, stacking and ARM\\
Renewal-process estimands & Mean, probability, log-hazard & Manual & Model-specific & Fitted distribution only & Parameters and predictions\\
\addlinespace
Bootstrap intervals after model selection & Percentile and Studentised bootstrap & Parameters and quantiles of one fit & Model-specific & Not provided & Analytic Wald intervals\\
Typical user effort for model-averaged renewal inference & Single workflow & Manual combination & Manual construction & Estimands re-derived by hand & Custom model objects required\\
\bottomrule
\end{tabular}
\end{table}

\section{Simulation study}\label{sec:simulation}

\subsection{Settings}\label{sec:simulation-design}

A simulation study was conducted to compare selected-model inference with AIC-weighted model averaging for estimating common renewal-process estimands. Specifically, it assessed whether model averaging improved point estimation or confidence-interval performance when the candidate models had varying levels of empirical support, and whether these gains came at the cost of reduced numerical stability. The quantities of interest were the mean inter-event time \(\mu = E(T)\), the event probability \(Pr(X \leq y)\), and the hazard function \(h(t)\) evaluated at selected time points.

Three simulation scenarios were considered. Simulation 1 represented a well-specified setting in which the true data-generating model was included among the \CRANpkg{marp} candidate models. Inter-event times were generated from a Gamma distribution,
\[
X \sim \mathrm{Gamma}(\alpha=3, \lambda = 0.01), 
\]
where \(\alpha\) is the shape parameter and \(\lambda\) is the rate parameter. The true mean was therefore
\[
\mu_{true}=\frac{\alpha}{\lambda}=300.
\]

The true time-to-event probability and hazard function were computed using the Gamma cumulative distribution and density functions.

Simulation 2 represented a challenging but still well-specified setting. Inter-event times were generated from a right-skewed Weibull distribution,
\[
X \sim \mathrm{Weibull}(k=0.7, \lambda=300),
\]
where \(k\) is the shape parameter and \(\lambda\) is the scale parameter. Because \(k<1\), this scenario has a decreasing hazard and a strongly right-skewed waiting-time distribution. Although the Weibull model is included in the candidate set, the strong skewness can make finite-sample model discrimination difficult, as several candidate models may provide similar empirical support. This scenario was therefore included as a challenging setting, particularly for upper-tail-sensitive quantities such as the mean inter-event time. The true mean was computed as
\[
\mu_{true}=\lambda \Gamma (1+\frac{1}{k}),
\]
with \(Pr(T \leq y)\) and \(h(t)\) computed using the Weibull distribution and density functions.

Simulation 3 represented a misspecified setting in which the true data-generating distribution was not included among the six \CRANpkg{marp} candidate models. Inter-event times were generated from the mixture,
\[
X \sim 0.7\,\mathrm{Gamma}(3, 0.01)
+ 0.3\,\mathrm{Weibull}(0.7, 500).
\]
The mixture density and distribution functions were calculated as
\[
f_{\text{mix}}(t)
= 0.7 f_{\text{Gamma}}(t) + 0.3 f_{\text{Weibull}}(t),
\]
and
\[
F_{\text{mix}}(t)
= 0.7 F_{\text{Gamma}}(t) + 0.3 F_{\text{Weibull}}(t).
\]

The corresponding true mean and event probability were computed from the
mixture distribution, and the true hazard was calculated as
\[
h_{\mathrm{mix}}(t)=\frac{f_{\mathrm{mix}}(t)}
{1-F_{\mathrm{mix}}(t)}.
\]
Because the mixture distribution was not among the six \CRANpkg{marp}
candidate models, generating-model results were excluded for this scenario.

For the main point-estimation simulation, sample sizes \(n=20,50,100\) were considered, with \(R=999\) independently simulated datasets for each scenario and sample size. The six candidate models were Poisson, gamma, log-logistic, Weibull, log-normal and Brownian passage time models. The fitting control parameter was set to \(m=10\). The \texttt{poisson\_rp()} model corresponds to the exponential waiting-time model associated with a homogeneous Poisson process.

For each dataset, \texttt{marp()} was used to obtain estimates from the minimum-AIC model and the AIC-weighted model-averaged estimator. Oracle estimates were additionally recorded for the well-specified gamma and Weibull scenarios by fitting the true data-generating family, with parameters estimated from each simulated dataset. The hazard function was evaluated at \(t=150\) and \(t=200\). The event probability was evaluated at \(y=304\), a threshold close to the mean inter-event time of 300 in the baseline gamma scenario. This provided a central probability target while retaining a common threshold across scenarios.

Confidence-interval performance was evaluated separately using \texttt{confint()}, because the bootstrap-based procedure was computationally intensive and could be numerically unstable in repeated simulations. The confidence-interval simulation used sample sizes \(n=50\) and \(n=100\) for each of the three scenarios, with \(R=999\) independently generated datasets for each scenario and sample size. Bootstrap parameters were set to \(B=999\) outer bootstrap resamples and \(BB=199\) inner bootstrap resamples, with \(\alpha=0.05\). Studentised bootstrap confidence intervals were assessed for the generating model, the minimum-AIC model and the AIC-weighted model-averaged estimator, where applicable. Generating-model confidence intervals were excluded for the misspecified mixture scenario.

\subsection{Performance measures and Monte Carlo uncertainty}\label{performance-measures-and-monte-carlo-uncertainty}

Point-estimation performance was summarised using empirical bias, empirical standard deviation across simulation replications, mean squared error (MSE) and root mean squared error (RMSE). Let \(\widehat{\psi}_r\) denote the estimate of a target quantity \(\psi\) in simulation replication \(r\), and let \(R_{\mathrm{finite}}\) denote the number of finite estimates available for that target, method and scenario. Empirical bias was calculated as
\[
\mathrm{Bias}(\widehat{\psi})
=
\frac{1}{R_{\mathrm{finite}}}
\sum_{r=1}^{R_{\mathrm{finite}}}
(\widehat{\psi}_r-\psi).
\]
The empirical standard deviation across simulation replications was calculated as
\[
\mathrm{SD}(\widehat{\psi})
=
\left[
\frac{1}{R_{\mathrm{finite}}-1}
\sum_{r=1}^{R_{\mathrm{finite}}}
\left\{
\widehat{\psi}_r
-
\frac{1}{R_{\mathrm{finite}}}\sum_{s=1}^{R_{\mathrm{finite}}}\widehat{\psi}_s
\right\}^2
\right]^{1/2}.
\]
The MSE was calculated as
\[
\mathrm{MSE}(\widehat{\psi})
=
\frac{1}{R_{\mathrm{finite}}}
\sum_{r=1}^{R_{\mathrm{finite}}}
(\widehat{\psi}_r-\psi)^2,
\]
and the RMSE was defined as
\[
\mathrm{RMSE}(\widehat{\psi})
=
\sqrt{\mathrm{MSE}(\widehat{\psi})}.
\]

Confidence-interval performance was summarised using empirical coverage, average interval width and failure rate. Coverage was calculated among successful confidence-interval computations as the proportion of valid intervals that contained the true target value. Average interval width was also calculated among successful confidence-interval computations. The failure rate was calculated as the proportion of simulation replications in which valid interval estimates were not returned.

Monte Carlo uncertainty was reported for selected simulation summaries. For conditional empirical coverage, the Monte Carlo standard error was estimated as
\[
\mathrm{MCSE}(\widehat{p})
=
\sqrt{
\frac{\widehat{p}(1-\widehat{p})}{R_{\mathrm{success}}}
},
\]
where \(\widehat{p}\) is the empirical coverage probability and \(R_{\mathrm{success}}\) is the number of valid confidence intervals. Approximate 95\% Wilson intervals were also calculated for empirical coverage.

For point-estimate bias, Monte Carlo uncertainty was estimated as
\[
\mathrm{MCSE}(\widehat{\mathrm{Bias}})
=
\frac{\mathrm{SD}(\widehat{\psi}-\psi)}
{\sqrt{R_{\mathrm{finite}}}}.
\]
Monte Carlo uncertainty for RMSE was also summarised using approximate intervals derived from the Monte Carlo variability of the squared errors. Confidence-interval failure rates and average interval widths were reported descriptively.

\subsection{Point estimation results}\label{point-estimation-results}

\begin{figure}[!htbp]

{\centering \includegraphics[width=1\linewidth,alt={Boxplots of replicate-level errors in mean inter-event-time estimates across three simulation scenarios, stratified by sample size and estimation method.}]{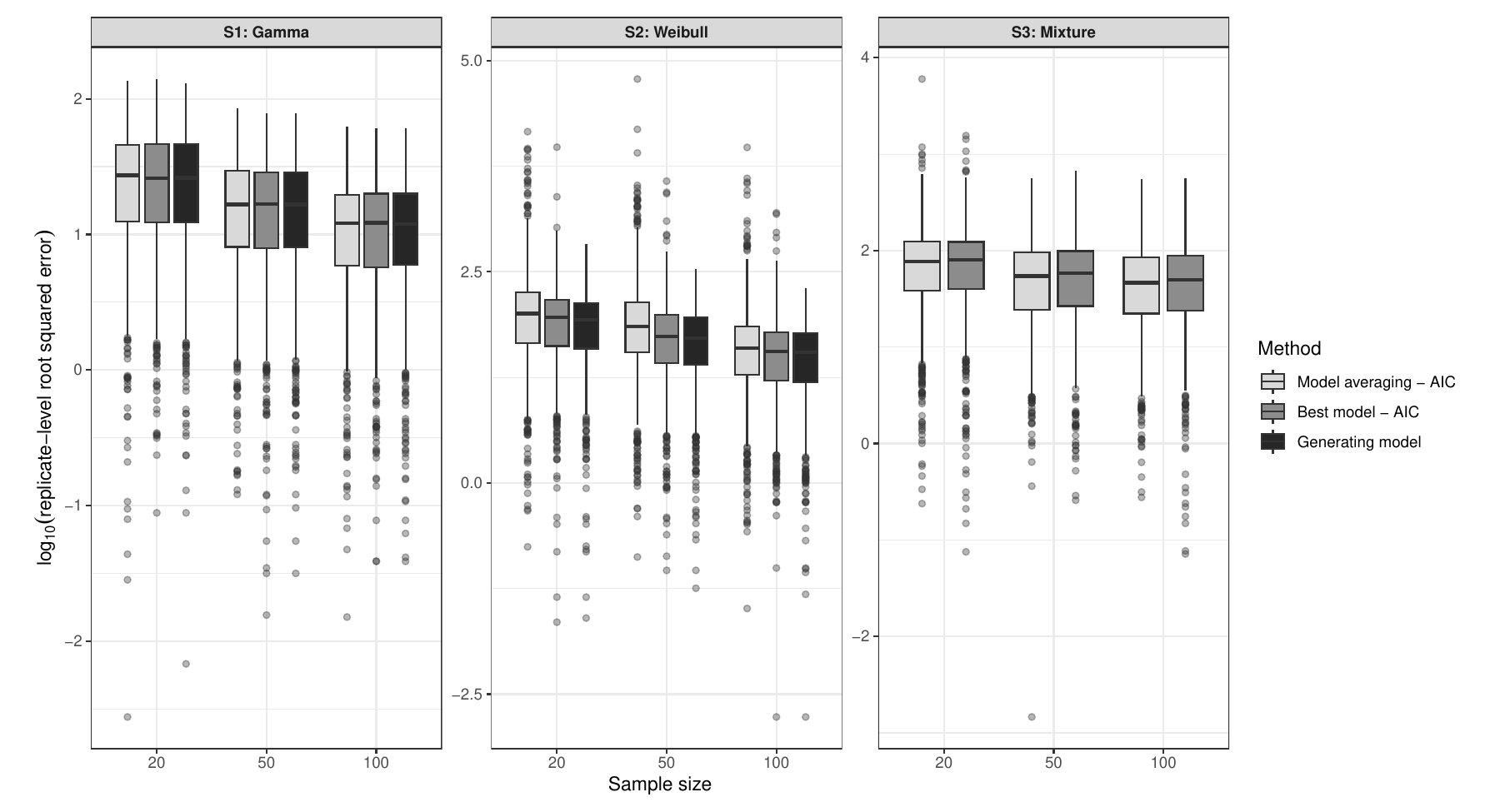} 

}

\caption{Distribution of replicate-level root squared errors for mean inter-event-time estimates across simulation scenarios. Values are shown on the log10 scale. Boxplots summarise the distribution across Monte Carlo replications for each sample size, method and scenario.}\label{fig:mean-replicate-error-boxplot}
\end{figure}

Figure \ref{fig:mean-replicate-error-boxplot} shows the distribution of replication-level errors for the mean inter-event-time estimates across the three simulation scenarios. Values are shown on the \(\log_{10}\) scale, where the plotted quantity is the root squared error for each replication, equivalent to the absolute estimation error. The boxplots summarise the empirical distribution of errors across Monte Carlo replications for each sample size, method and scenario.

In the Gamma scenario, Model averaging--AIC and Best model--AIC showed similar error distributions for the mean inter-event time. Errors generally decreased as sample size increased. The generating-model estimator, included as an oracle benchmark, showed similar performance in this well-specified setting.

In the Weibull scenario, errors were generally larger than in the Gamma scenario, reflecting the stronger right-skewness and decreasing-hazard structure of the data-generating distribution. Across all sample sizes, the generating-model estimator tended to have the lowest median error, as expected for an oracle benchmark. Best model--AIC was close to the generating-model estimator, while Model averaging--AIC had slightly higher median errors and a somewhat wider upper tail, particularly at \(n=20\) and \(n=50\). These differences became smaller at \(n=100\).

In the Mixture scenario, no generating-model estimator was reported because the true mixture distribution was not included among the candidate models. At \(n=20\), Model averaging--AIC and Best model--AIC showed similar median errors, although Model averaging--AIC had slightly larger upper-tail errors. At \(n=50\) and \(n=100\), the two methods remained broadly comparable, with Model averaging--AIC showing slightly lower median errors and Best model--AIC showing slightly smaller upper-tail errors in some settings.

Overall, the point-estimation results suggest that the relative performance of model averaging and minimum-AIC model selection depends on the data-generating scenario. Model averaging did not uniformly dominate selected-model estimation, but it provided comparable mean-estimation performance in the well-specified scenarios considered here while retaining the advantage of incorporating model-selection uncertainty. The generating-model estimator is included only for the Gamma and Weibull scenarios as an oracle benchmark, because in applied analyses the true data-generating family would not be known.

Additional point-estimation summaries for the mean inter-event time, including \(\log_{10}(\mathrm{RMSE})\) across scenarios, sample sizes and methods, are provided in the Supplementary Material, Section 2.1.

\subsection{Interval-estimation results}\label{interval-estimation-results}

Confidence-interval performance was evaluated using \texttt{confint()} with \(B=999\) outer bootstrap resamples and \(BB=199\) inner bootstrap resamples. Table \ref{tab:ci-failure-pdf} summarises the failure rate of the Studentised bootstrap confidence-interval procedure across scenarios and sample sizes.

\begin{table}[!htbp]
\centering
\caption{\label{tab:ci-failure-pdf}Failure rate of Studentised bootstrap confidence-interval calculations across simulation scenarios. The failure rate was calculated as the proportion of simulation replications in which confint() did not return valid confidence-interval estimates.}
\centering
\fontsize{8}{10}\selectfont
\begin{tabular}[t]{lrlr}
\toprule
Scenario & Sample size & CI failures / total & CI failure rate\\
\midrule
Gamma well-specified & 50 & 0 / 999 & 0.000\\
Gamma well-specified & 100 & 0 / 999 & 0.000\\
Weibull challenging & 50 & 93 / 999 & 0.093\\
Weibull challenging & 100 & 93 / 999 & 0.093\\
Mixture misspecified & 50 & 15 / 999 & 0.015\\
\addlinespace
Mixture misspecified & 100 & 24 / 999 & 0.024\\
\bottomrule
\end{tabular}
\end{table}

Confidence-interval computation was fully successful in the well-specified Gamma scenario, with no failures at either sample size. Failures occurred most often in the challenging Weibull scenario, with failure rates of 9.3\% at both \(n=50\) and \(n=100\). The Mixture scenario had lower failure rates of 1.5\% at \(n=50\) and 2.4\% at \(n=100\). In the final simulation run, failed computations were recorded as invalid confidence-interval bounds. These results indicate that the Studentised bootstrap procedure was generally feasible, but that numerical issues remained in more challenging settings.

\begin{figure}[H]

{\centering \includegraphics[width=1\linewidth,alt={Point and error-bar plot of empirical confidence-interval coverage across simulation scenarios, target quantities and sample sizes. Error bars show Monte Carlo standard errors, and a dashed horizontal reference line marks 95\% coverage.}]{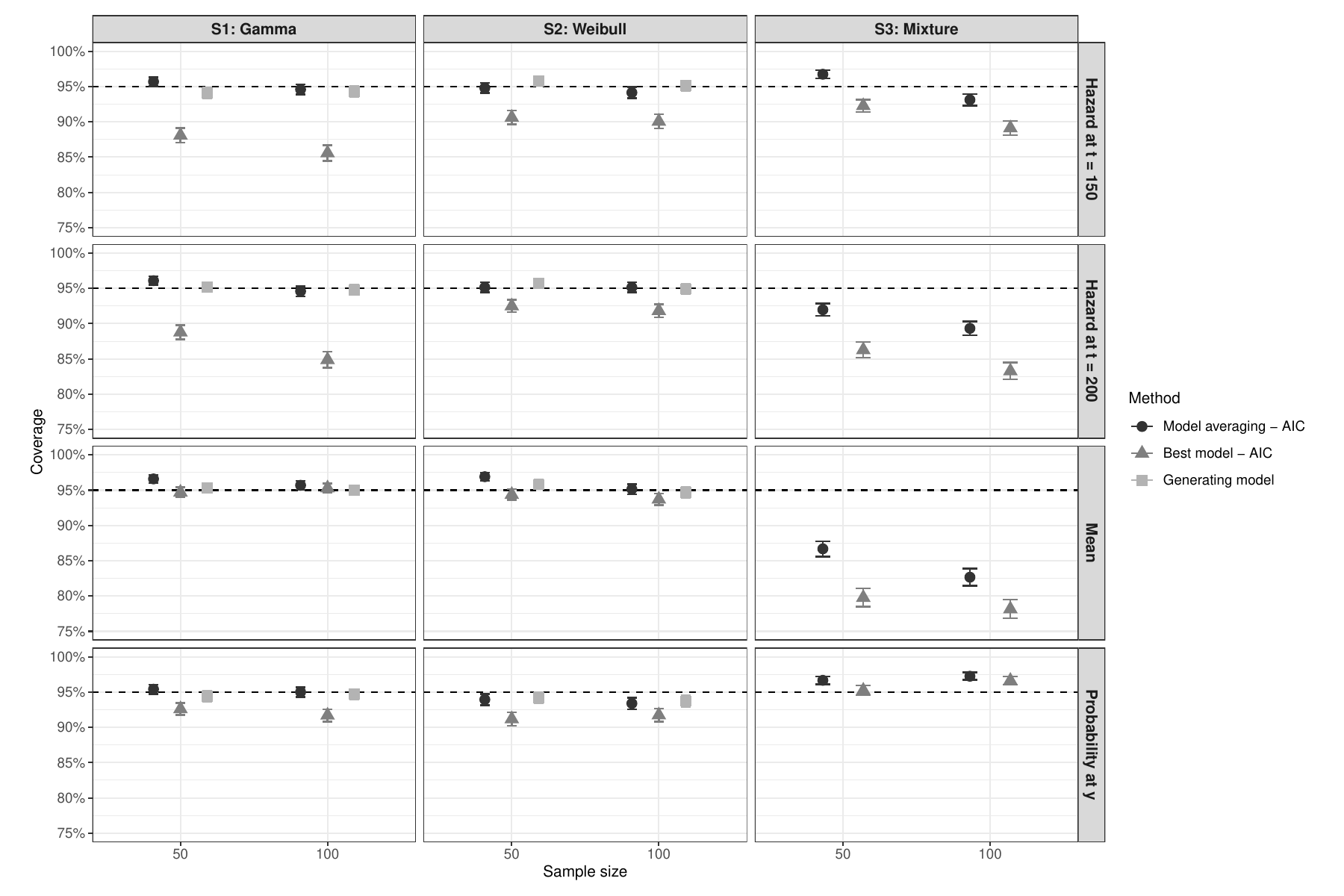} 

}

\caption{Empirical coverage of Studentised bootstrap confidence intervals across simulation scenarios, target quantities and sample sizes. Coverage is calculated among successful confidence-interval computations, so the reported coverage values are conditional on successful interval construction. Error bars show one Monte Carlo standard error for empirical coverage. The dashed horizontal line indicates nominal 95\% coverage. The y-axis is zoomed to 75--100\% to highlight differences around nominal coverage.}\label{fig:ci-coverage}
\end{figure}

Empirical coverage among successful confidence-interval computations is shown in Figure \ref{fig:ci-coverage}. Coverage was assessed for the mean inter-event time, event probability \(\Pr(X\leq y)\), and hazards at \(t=150\) and \(t=200\). The dashed horizontal line indicates the nominal 95\% coverage level.

In the Gamma scenario, Model averaging--AIC generally achieved conditional coverage close to or slightly above the nominal level across target quantities. Best model--AIC showed lower coverage for the hazard estimands, particularly at \(t=150\) and \(t=200\), where coverage was below 90\% in some settings. For the mean and event-probability estimands, the differences between Model averaging--AIC and Best model--AIC were smaller. The generating-model intervals, included as an oracle benchmark, also had coverage close to the nominal level in this well-specified setting.

In the Weibull scenario, Model averaging--AIC and the generating-model estimator generally achieved conditional coverage close to the nominal level among successful confidence-interval computations. Best model--AIC again showed lower coverage for the hazard estimands, while differences between methods were smaller for the mean and event-probability estimands. These results should be interpreted together with the failure rates in Table \ref{tab:ci-failure-pdf} because the Weibull scenario had the highest rate of unsuccessful confidence-interval computations.

In the Mixture scenario, coverage was more variable across target quantities. Model averaging--AIC had higher conditional coverage than Best model--AIC for the mean and hazard estimands, although coverage for the mean remained below the nominal level. For the event-probability estimand, both methods achieved coverage close to or above the nominal level, with Model averaging--AIC slightly above nominal at both sample sizes.

Overall, the interval-estimation results suggest that AIC-weighted model averaging can improve or maintain conditional coverage relative to minimum-AIC selection among successful confidence-interval computations, particularly for hazard estimands. However, these comparisons should be interpreted alongside Monte Carlo uncertainty and confidence-interval failure rates. Failed confidence-interval computations are part of the practical operating characteristics of the method.

Additional confidence-interval summaries, including conditional coverage ranges, mean coverage and failure rates across target quantities, are provided in the Supplementary Material, Section 2.2.

\subsection{Numerical stability}\label{numerical-stability}

For point estimation, no non-finite estimates were observed in the simulation results, although the Weibull scenario remained the most challenging setting, with larger finite replicate-level errors than in the Gamma scenario. For interval estimation, failed Studentised bootstrap confidence-interval computations occurred in the Weibull and Mixture scenarios but not in the Gamma scenario. The highest failure rate was 9.3\% in the Weibull scenario, and all recorded failures in the final simulation run were due to invalid confidence-interval bounds. These results indicate that numerical stability should be reported alongside estimation and coverage summaries, particularly for bootstrap procedures involving repeated model fitting and model averaging.

\section{Applications}\label{sec:applications}

We illustrate the use of \CRANpkg{marp} using two real time-to-event datasets. The first is a biomedical recurrence example based on seizure onset times from the CHB-MIT Scalp EEG Database. The second is an engineering reliability example based on airplane glass failure times. These examples were chosen to demonstrate how \CRANpkg{marp} can be used when the number of observed waiting times is modest and when several candidate parametric models provide plausible descriptions of the data.
For readability, the code shown in this section focuses on the core package workflow; data preprocessing, plotting details and saved-output generation are provided in the accompanying reproducibility scripts.

\subsection{Seizure recurrence data}\label{sec:seizure}

\begin{figure}[!htbp]

{\centering \includegraphics[width=0.7\linewidth,alt={Histogram of 13 preprocessed inter-seizure times for subject chb-15, showing the distribution of waiting times used as input to the renewal-process models.}]{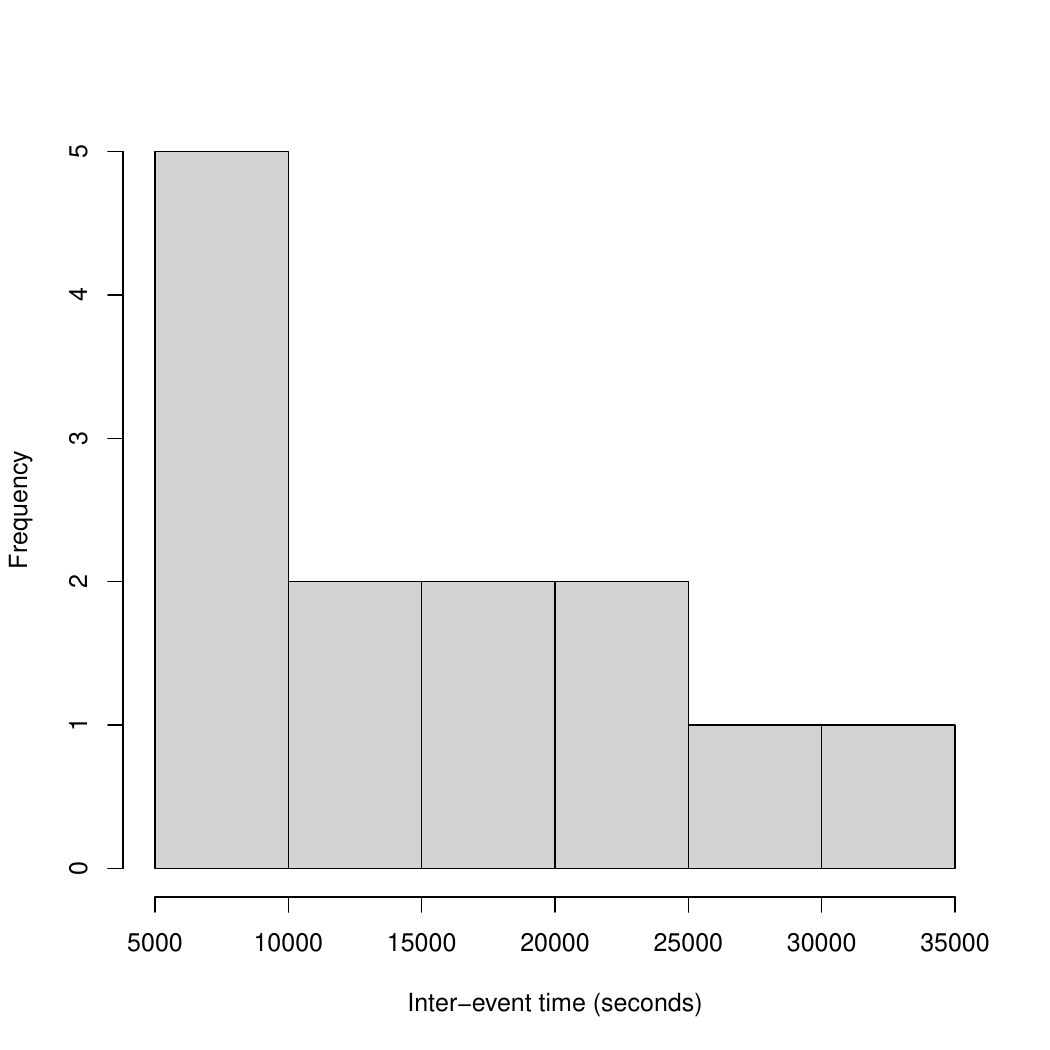} 

}

\caption{Histogram of the preprocessed inter-seizure times for subject chb-15 from the CHB-MIT Scalp EEG Database. Seizure events separated by less than 1800 seconds were treated as clustered events before constructing the inter-event times. The resulting 13 inter-seizure times form the input to the renewal models fitted using \CRANpkg{marp}.}\label{fig:hist-seizure}
\end{figure}

\textbf{Data.}
The CHB-MIT Scalp EEG Database contains scalp electroencephalography recordings from paediatric subjects with intractable seizures, collected at Children's Hospital Boston \citep{guttag2010chbmit}. In addition to the EEG recordings, seizure onset and end times are available for the recorded seizure events. In this application, the raw EEG signals were not modelled directly. Instead, annotated seizure onset times were used to construct inter-seizure times.
\textbf{Approach.}
To make the data more suitable for renewal-process modelling, the analysis was restricted to a single subject, \texttt{chb-15}. This avoids pooling seizure times across different subjects, which would not be consistent with the independent and identically distributed inter-event-time assumption. A preprocessing step was also applied to reduce dependence caused by short seizure clusters: seizure events separated by less than 1800 seconds were treated as clustered events before constructing the inter-event times. After preprocessing, 13 inter-seizure times remained for analysis. The distribution of these inter-seizure times is shown in Figure \ref{fig:hist-seizure}.

\textbf{Results.}
The six candidate renewal models were fitted to the preprocessed inter-seizure times using \CRANpkg{marp}. Model-averaged uncertainty was assessed using \texttt{confint()} with \(B=999\) outer bootstrap resamples and \(BB=999\) inner bootstrap resamples. Table \ref{tab:seizure-model-comparison-pdf} reports the AIC, BIC and AIC weights for the fitted models. The BPT model had the smallest AIC and BIC, although the log-normal and gamma models also received non-negligible AIC weights. This indicates that selected-model and model-averaged summaries are related but not identical.

\FloatBarrier

\begin{table}[H]
\centering
\caption{\label{tab:seizure-model-comparison-pdf}Model support for the seizure recurrence data from subject chb-15. The table reports AIC, BIC and AIC weights for the six fitted candidate renewal models.}
\centering
\fontsize{8}{10}\selectfont
\begin{tabular}[t]{lrrr}
\toprule
Model & AIC & BIC & AIC weight\\
\midrule
BPT & 271.11 & 272.24 & 0.315\\
Log-normal & 271.53 & 272.66 & 0.255\\
Gamma & 272.24 & 273.37 & 0.179\\
Log-logistic & 272.89 & 274.02 & 0.129\\
Weibull & 273.10 & 274.23 & 0.116\\
\addlinespace
Poisson & 279.07 & 279.63 & 0.006\\
\bottomrule
\end{tabular}
\end{table}

\begin{figure}[!htbp]

{\centering \includegraphics[width=1\linewidth,alt={Two-panel figure for subject chb-15. Panel (a) shows AIC weights across six candidate renewal models. Panel (b) shows the model-averaged log-hazard estimate across inter-seizure times with a shaded confidence interval.}]{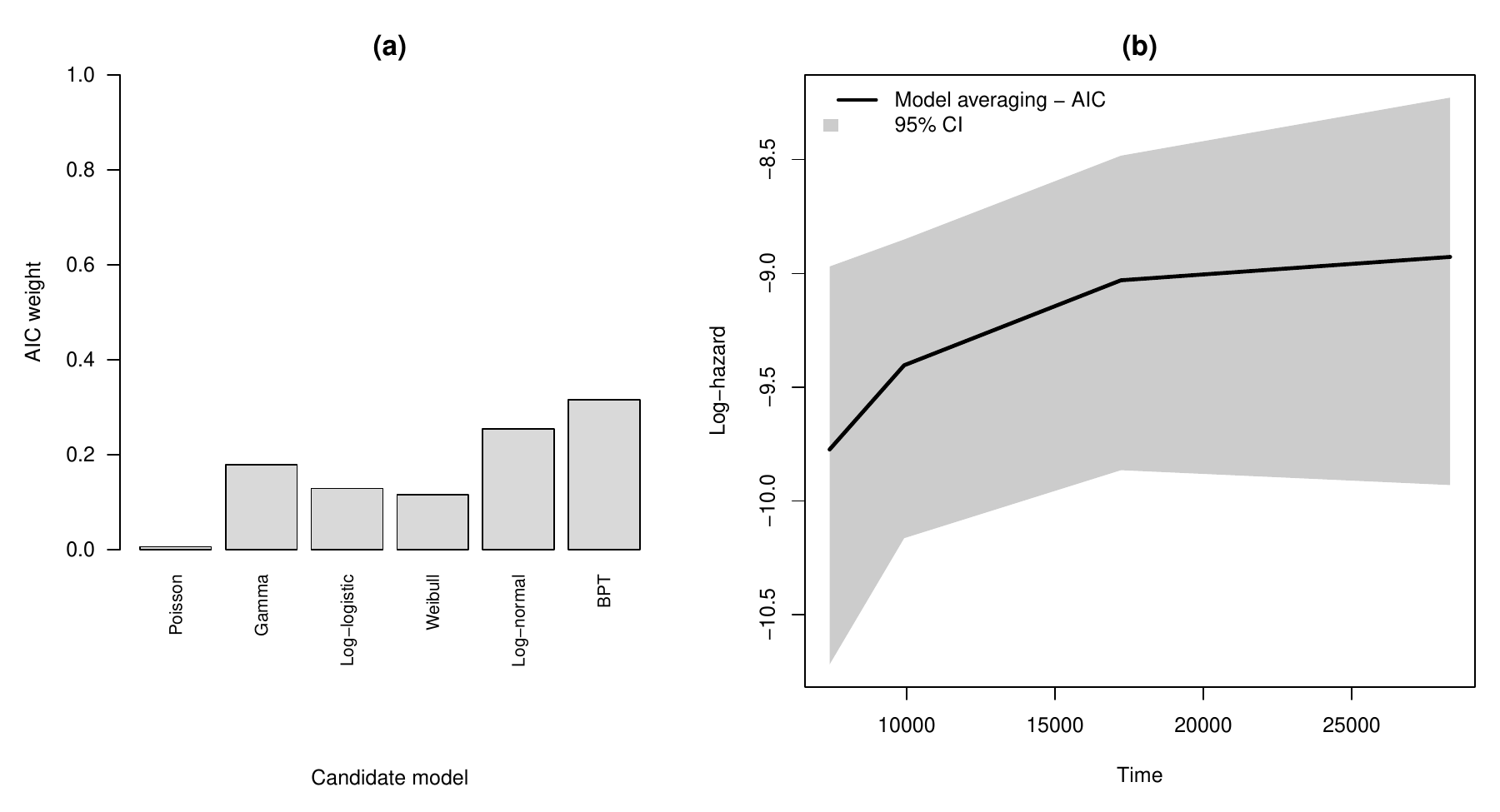} 

}

\caption{Model support and model-averaged log-hazard estimates for the seizure recurrence data from subject chb-15. Panel (a) shows the AIC weights for the six fitted candidate renewal models. Panel (b) shows the AIC-weighted model-averaged log-hazard estimate, with the shaded region indicating the model-averaged confidence interval.}\label{fig:hazard-seizure}
\end{figure}

The core application workflow is shown below. The object \texttt{dat} denotes the preprocessed inter-seizure times. The fitted object supplies the model-support summaries in Table \ref{tab:seizure-model-comparison-pdf}, while the fitted and confidence-interval objects supply the log-hazard estimates and interval limits in Figure \ref{fig:hazard-seizure}.

\begin{verbatim}
t_eval <- as.numeric(
  quantile(dat, probs = seq(0.1, 0.9, length.out = 4))
)
y_eval <- mean(dat)
m <- 80

fit_seizure <- marp::marp(
  data = dat,
  t = t_eval,
  m = m,
  y = y_eval
)
reference_model <- fit_seizure$model_best

ci_seizure <- confint(
  fit_seizure,
  data = dat,
  B = 999,
  BB = 999,
  level = 0.95
)

model_table <- data.frame(
  model = c(
    "Poisson", "Gamma", "Log-logistic",
    "Weibull", "Log-normal", "BPT"
  ),
  AIC = fit_seizure$AIC,
  BIC = fit_seizure$BIC,
  weight_AIC = fit_seizure$weights_AIC
)

hazard_table <- data.frame(
  time = t_eval,
  log_hazard = fit_seizure$haz_aic,
  lower = ci_seizure$student_CI$haz_lower_ma,
  upper = ci_seizure$student_CI$haz_upper_ma
)
\end{verbatim}

Figure \ref{fig:hazard-seizure} summarises the model support and model-averaged log-hazard estimates. Panel (a) shows that model support was distributed across several candidate models, with the BPT, log-normal and gamma models receiving the largest AIC weights. Panel (b) shows the AIC-weighted model-averaged log-hazard estimate over selected empirical quantiles of the inter-seizure-time distribution. The estimated log-hazard increased over the observed range, although the confidence interval was wide, reflecting the small number of inter-event times.

This example illustrates how \CRANpkg{marp} can be used in a small-sample biomedical recurrence setting. Because only 13 inter-event times were available, the model weights and bootstrap interval should be interpreted cautiously. The main role of model averaging here is to avoid conditioning inference entirely on a single selected renewal distribution.

\FloatBarrier

\subsection{Airplane glass failure data}\label{sec:glass}

\textbf{Data.}
The second application uses the airplane glass failure-time data from the Wolfram Data Repository \citep{wolframairplaneglass}. The data originate from a fracture-mechanics study of glass aircraft windows by Fuller et al. \citep{fuller1994fracture}. The dataset contains 31 observations measuring the time to rupture for airplane glass specimens subjected to increasing pressure.

\begin{figure}[!htbp]

{\centering \includegraphics[width=0.75\linewidth,alt={Histogram of the Fuller airplane-glass failure times, showing the distribution of observed engineering failure-time measurements.}]{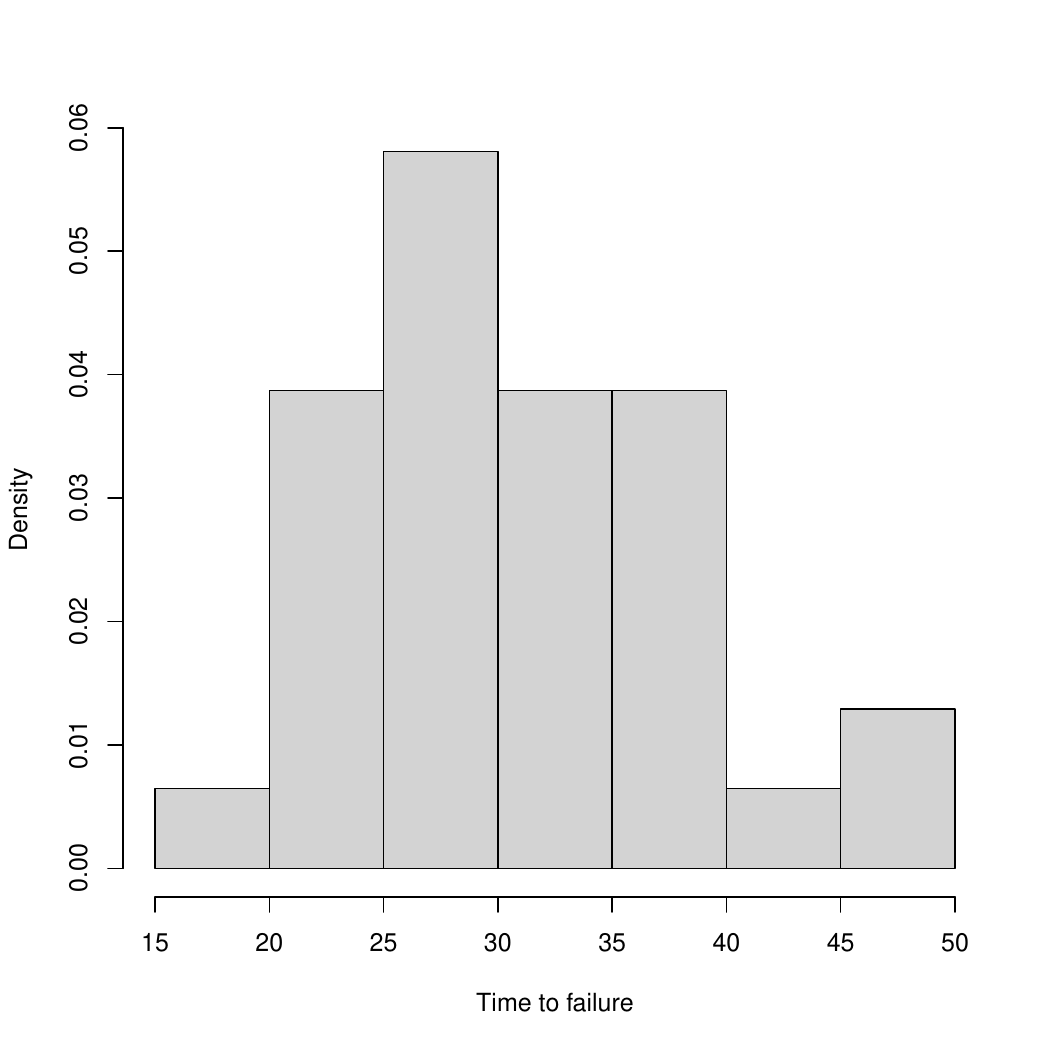} 

}

\caption{Histogram of the Fuller airplane-glass failure times. The observations are independent failure-time measurements from glass aircraft-window specimens and are analysed as an engineering reliability illustration of the parametric waiting-time models implemented in \CRANpkg{marp}.}\label{fig:hist-fuller}
\end{figure}

Unlike the seizure data, these observations are not recurrent events from a single subject. They are independent failure-time observations from multiple specimens. The data are therefore analysed as an illustration of parametric waiting-time model averaging for independent failure-time observations, rather than as a renewal-process example. The distribution of observed failure times is shown in Figure \ref{fig:hist-fuller}.

\textbf{Results.}
The six candidate waiting-time models were fitted to the airplane glass failure-time data using \CRANpkg{marp}. Model-averaged uncertainty was assessed using \texttt{confint()} with \(B=999\) outer bootstrap resamples and \(BB=999\) inner bootstrap resamples. Table \ref{tab:glass-model-comparison-pdf} reports the AIC, BIC and AIC weights for the fitted models. The BPT model had the smallest AIC and BIC, but the log-normal and gamma models had similar AIC values and non-negligible AIC weights. This provides a useful example where the selected model and the model-averaged estimate are not the same summary of the fitted candidate set.

\FloatBarrier

\begin{table}[H]
\centering
\caption{\label{tab:glass-model-comparison-pdf}Model support for the airplane glass failure-time data. The table reports AIC, BIC and AIC weights for the six fitted candidate waiting-time models.}
\centering
\fontsize{8}{10}\selectfont
\begin{tabular}[t]{lrrr}
\toprule
Model & AIC & BIC & AIC weight\\
\midrule
BPT & 211.90 & 214.77 & 0.301\\
Log-normal & 212.02 & 214.89 & 0.284\\
Gamma & 212.23 & 215.10 & 0.256\\
Log-logistic & 214.23 & 217.10 & 0.094\\
Weibull & 214.98 & 217.85 & 0.065\\
\addlinespace
Poisson & 276.53 & 277.96 & 0.000\\
\bottomrule
\end{tabular}
\end{table}

The code structure is the same as in the seizure application, with a denser grid of evaluation times because more observations were available. Here \texttt{dat} denotes the vector of airplane glass failure times.

\begin{verbatim}
t_eval <- as.numeric(
  quantile(dat, probs = seq(0.1, 0.9, length.out = 7))
)
y_eval <- mean(dat)
m <- 100

fit_glass <- marp::marp(
  data = dat,
  t = t_eval,
  m = m,
  y = y_eval
)
fit_glass
summary(fit_glass)

ci_glass <- confint(
  fit_glass,
  data = dat,
  B = 999,
  BB = 999,
  level = 0.95
)
\end{verbatim}

\begin{figure}[!htbp]

{\centering \includegraphics[width=1\linewidth,alt={Two-panel figure for the airplane glass failure-time data. Panel (a) shows AIC weights for six fitted candidate waiting-time models. Panel (b) shows the model-averaged log-hazard estimate across failure times with a shaded confidence interval.}]{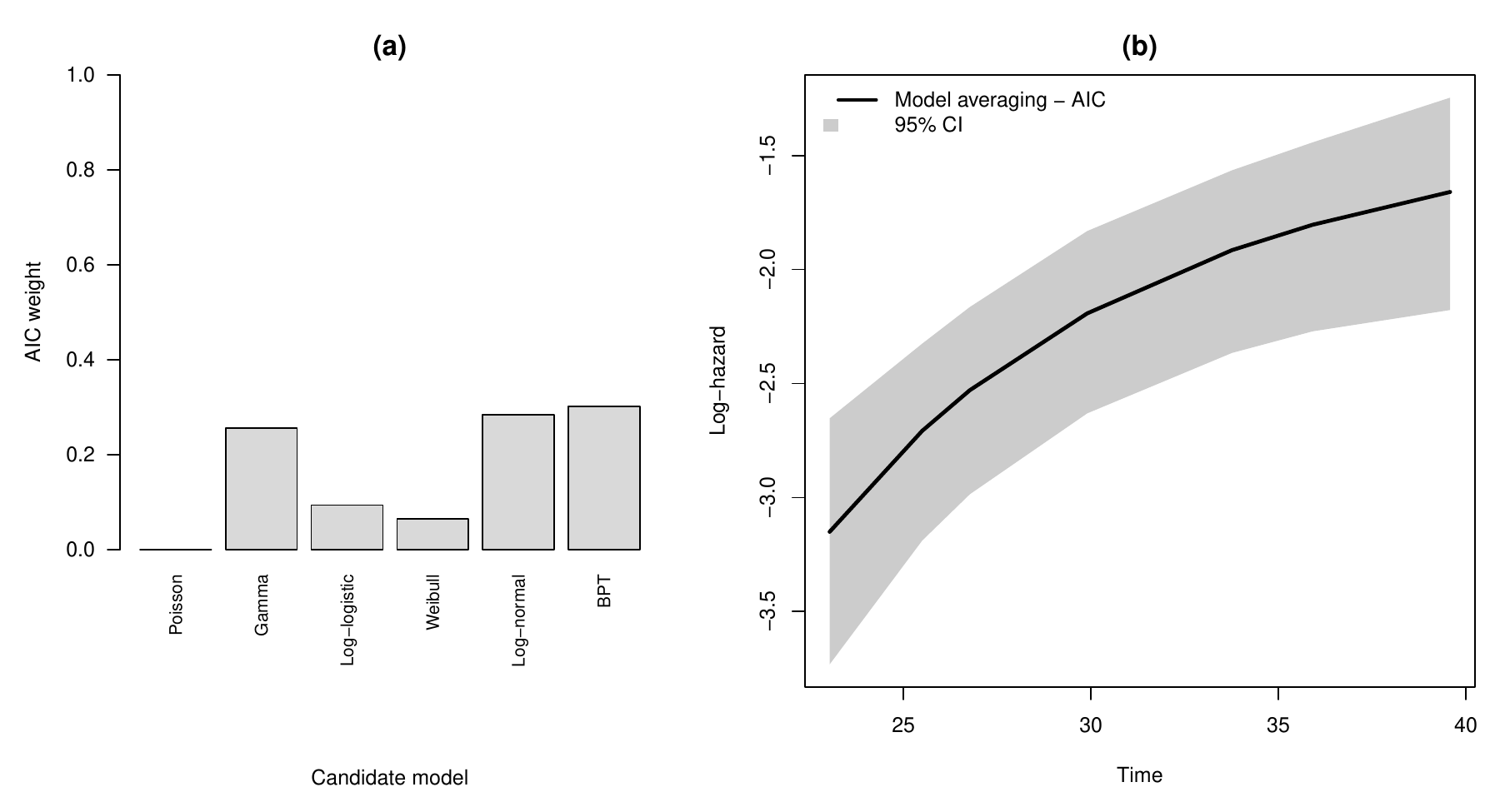} 

}

\caption{Model support and model-averaged log-hazard estimates for the airplane glass failure-time data. Panel (a) shows the AIC weights for the six fitted candidate waiting-time models. Panel (b) shows the AIC-weighted model-averaged log-hazard estimate, with the shaded region indicating the model-averaged confidence interval.}\label{fig:hazard-fuller}
\end{figure}

Figure \ref{fig:hazard-fuller} summarises the main model-averaging results. Panel (a) shows that model support was distributed across several flexible two-parameter models, with the BPT, log-normal and gamma models receiving the largest AIC weights. Panel (b) shows the AIC-weighted model-averaged log-hazard estimate over selected failure-time quantiles, together with the model-averaged confidence interval. The estimated log-hazard increased over the observed range of failure times, while the confidence interval reflects uncertainty in the model-averaged hazard estimate.

This example illustrates the use of \CRANpkg{marp} when several parametric failure-time distributions receive comparable empirical support. In this setting, model averaging provides a way to summarise the hazard trend while reflecting uncertainty across the candidate waiting-time models.

Additional application summaries, including model-comparison results, numerical log-hazard estimates and diagnostic comparisons with empirical log-hazard curves for both datasets, are provided in Supplementary Material, Section 3.

\section{Discussion}\label{sec:discussion}

\subsection{Main findings}\label{main-findings}

This paper introduced \CRANpkg{marp}, an \texttt{R} package for model comparison and model-averaged inference across parametric renewal-process models. The package provides a unified workflow for fitting candidate waiting-time distributions, comparing models using information criteria, constructing selected-model and model-averaged point estimates, and obtaining bootstrap confidence intervals. Its main contribution is computational: it makes model-averaged renewal-process inference easier to implement, report and reproduce.

The simulation study showed that the performance of model averaging depends on both the target functional and the candidate model set. In the well-specified gamma scenario, selected-model and model-averaged estimators performed similarly. In the challenging Weibull scenario, mean-estimation errors were larger overall, reflecting the difficulty of estimating upper-tail-sensitive functionals under strongly right-skewed waiting-time distributions. Model averaging did not uniformly improve point estimation, but it provided a useful way to represent uncertainty across the candidate model set. For interval estimation, AIC-weighted model averaging generally achieved conditional coverage close to the nominal level among successful computations and often improved or maintained coverage relative to minimum-AIC selection, particularly for hazard estimands. However, bootstrap failures occurred in the Weibull and mixture scenarios, indicating that coverage should be interpreted alongside failure rates and operational performance.

The two applications illustrated how \CRANpkg{marp} can be used in modest-sized time-to-event datasets. In both applications, BPT was selected by AIC and BIC, but the AIC weights were not concentrated entirely on that model. This shows that selected-model and model-averaged summaries are related but not identical.

\subsection{Practical implications}\label{practical-implications}

The main practical value of model averaging is that it makes model-selection uncertainty visible. Selected-model inference conditions on a single chosen distribution, whereas AIC-weighted model averaging combines fitted quantities across candidate models according to their relative empirical support \citep{buckland1997model, burnham2004multimodel}. This is useful for renewal-process data because different waiting-time distributions imply different hazard shapes. When sample sizes are modest, the data may not clearly distinguish between these shapes, so reporting model weights, selected-model estimates and model-averaged estimates gives a clearer account of sensitivity to the assumed waiting-time distribution.

The integrated workflow in \CRANpkg{marp} also reduces the amount of manual implementation required. Without a dedicated package, users would need to fit each candidate model separately, calculate information criteria and model weights, combine model-specific estimates, and repeat these steps during bootstrap resampling. \CRANpkg{marp} provides these steps in a common framework, supporting more transparent and reproducible applied analyses.

\subsection{Limitations and future work}\label{limitations-and-future-work}

Several limitations should be noted. First, model averaging does not remove the need for careful model checking. If all candidate models are poor descriptions of the data, the model-averaged estimate may still be biased or misleading. AIC weights and model-averaged estimates should therefore be interpreted alongside graphical diagnostics, subject-matter knowledge and sensitivity analyses.

Second, \CRANpkg{marp} currently focuses on strict renewal processes, where the inter-event times are assumed to be independent and identically distributed. This assumption excludes settings with censoring, covariates, subject-level heterogeneity or dependence between successive waiting times. Future development could extend \CRANpkg{marp} beyond this setting to regression-based recurrent-event models, frailty models and other frameworks that allow covariates, subject-level heterogeneity or dependence between event times \citep{cook2007statistical, lawless1987regression, hougaard2000analysis}.

Third, bootstrap uncertainty estimation can be computationally expensive and numerically unstable in challenging settings. In the simulation study, Studentised bootstrap failures occurred in the Weibull and mixture scenarios, with the highest failure rate in the Weibull scenario. These failures indicate that bootstrap interval performance should be assessed not only through conditional coverage, but also through failure rates and numerical diagnostics. In applied analyses, users should inspect model-specific fitted values, model weights, bootstrap selection frequencies and failure rates. Larger studies may also require parallelisation or careful choices of the number of outer and inner bootstrap resamples \citep{efron1993bootstrap}.

Further extensions of \CRANpkg{marp} could include additional waiting-time distributions, more flexible semi-parametric alternatives and goodness-of-fit tools. Bayesian model averaging could be incorporated as an alternative to information-criterion weighting, allowing prior information and posterior model uncertainty to be represented more explicitly \citep{hoeting1999bayesian}. Bootstrap-based model averaging strategies could also be explored as another way to assess the stability of model weights and model-averaged estimates \citep{song2024bootstrap}. Automated reporting functions for model-comparison tables, model-averaged estimates and diagnostic plots would further support reproducible analysis.

\section{Software and data availability}\label{sec:software-data}

The \CRANpkg{marp} package version 0.1.1, the version relevant to this article, is available from CRAN at \url{https://CRAN.R-project.org/package=marp}. The package source code is also available from GitHub at \url{https://github.com/kanji709/marp} and is distributed under the MIT licence. Scripts and saved results used to reproduce the simulation study, application analyses, tables and figures are supplied with the accompanying supplementary materials. The CHB-MIT Scalp EEG Database is publicly available from PhysioNet, and the airplane glass failure-time data are publicly available from the Wolfram Data Repository.

\section{Conclusion}\label{sec:conclusion}

This paper introduced \CRANpkg{marp}, an open-source \texttt{R} package for model comparison and model-averaged inference across parametric waiting-time distributions. By integrating model fitting, information-criterion comparison, selected-model and model-averaged estimation, and bootstrap uncertainty quantification, \CRANpkg{marp} provides a reproducible workflow for renewal-process analyses.

The main value of \CRANpkg{marp} is that it makes distributional uncertainty easier to assess and report in applied renewal-process modelling. Users can compare selected-model and model-averaged summaries, inspect model weights, and quantify uncertainty using bootstrap intervals within a single framework. The simulation study and applications show that model averaging can be useful, but should be accompanied by diagnostics, model-weight summaries and bootstrap failure rates. Future development will focus on stronger diagnostic tools, more robust optimisation, additional candidate distributions, covariate-dependent models, and extensions for heterogeneous or dependent recurrent-event processes.

\section{Funding}\label{funding}

JK is supported by Charles Perkins Centre's Jennie Mackenzie Research Fund.

\section{Conflict of interest}\label{conflict-of-interest}

The authors declare no conflicts of interest.
Correspondence concerning this article should be addressed to Jie Kang at \href{mailto:jie.kang@sydney.edu.au}{\nolinkurl{jie.kang@sydney.edu.au}}.

\clearpage
\bibliographystyle{abbrvnat}
\bibliography{RJreferences.bib}

\clearpage
\newgeometry{margin=1in}
\setlength{\parindent}{0pt}
\setlength{\parskip}{6pt plus 2pt minus 1pt}
\renewcommand{\textbf}[1]{#1}
\setcounter{section}{0}
\setcounter{subsection}{0}
\setcounter{figure}{0}
\setcounter{table}{0}
\renewcommand{\theHsection}{S\arabic{section}}
\renewcommand{\theHsubsection}{S\arabic{section}.\arabic{subsection}}
\renewcommand{\theHfigure}{S\arabic{figure}}
\renewcommand{\theHtable}{S\arabic{table}}
\begin{center}
{\LARGE Supplementary Material\par}
\vspace{0.6em}
{\large marp: Model-Averaged Inference for Parametric Renewal Processes in R\par}
\end{center}
\vspace{1.5em}

\hypertarget{package-functions-and-returned-objects}{%
\section{Package functions and returned
objects}\label{package-functions-and-returned-objects}}

This supplementary material summarises the main user-facing functions in
\texttt{marp} and the principal objects returned by each function.

\hypertarget{model-specific-fitting-functions}{%
\subsection{Model-specific fitting
functions}\label{model-specific-fitting-functions}}

The six model-specific fitting functions are provided in Table S1:

\textbf{Table S1: Model-specific fitting functions implemented in
\texttt{marp}.}

\begin{center}
\small

\begin{tabular}{ll}
\toprule
Model & Function\\
\midrule
Poisson (Exponential) & poisson\_rp()\\
Gamma & gamma\_rp()\\
Log-logistic & loglogis\_rp()\\
Weibull & weibull\_rp()\\
Log-normal & lognorm\_rp()\\
\addlinespace
Brownian passage time & bpt\_rp()\\
\bottomrule
\end{tabular}
\end{center}

Each function fits a single renewal model to a vector of observed
inter-event times. The common arguments are summarised in Table S2.

\textbf{Table S2: Common arguments for the model-specific fitting
functions.}

\begin{center}
\small

\begin{tabular}{ll}
\toprule
Argument & Description\\
\midrule
data & An (n)-dimensional vector of observed inter-event times.\\
t & Time point or vector of time points at which the hazard function is evaluated.\\
y & Cut-off time used to estimate the event probability $\Pr(X \leq y)$.\\
\bottomrule
\end{tabular}
\end{center}

The principal objects returned by the model-specific fitting functions
are summarised in Table S3.

\textbf{Table S3: Main objects returned by the model-specific fitting
functions.}

\begin{center}
\small

\begin{tabular}{ll}
\toprule
Object & Description\\
\midrule
par1 & First estimated model parameter.\\
par2 & Second estimated model parameter, where applicable.\\
logL & Maximised log-likelihood for the fitted candidate model.\\
AIC & Akaike information criterion.\\
BIC & Bayesian information criterion.\\
\addlinespace
mu\_hat & Estimated mean inter-event time.\\
pr\_hat & Estimated event probability, reported on the logit scale.\\
haz\_hat & Estimated hazard, reported on the log scale.\\
\bottomrule
\end{tabular}
\end{center}

\hypertarget{model-selection-and-model-averaging}{%
\subsection{Model selection and model
averaging}\label{model-selection-and-model-averaging}}

The main model-averaging function is:

\texttt{marp(data,\ t,\ m,\ y,\ which.model)}

This function fits all six candidate models, constructs
information-criterion weights and returns selected-model and
model-averaged estimates. In addition to the common arguments in Table
S2, \texttt{marp()} uses the arguments shown in Table S4.

\textbf{Table S4: Additional arguments used by \texttt{marp()}.}

\begin{center}
\small

\begin{tabular}{ll}
\toprule
Argument & Description\\
\midrule
m & Number of repeated random-start optimisations.\\
which.model & Generating or reference model, when known.\\
\bottomrule
\end{tabular}
\end{center}

The main model-selection and model-averaging outputs returned by
\texttt{marp()} are summarised in Table S5.

\textbf{Table S5: Main model-selection and model-averaging objects
returned by \texttt{marp()}.}

\begin{center}
\small

\begin{tabular}{ll}
\toprule
Object & Description\\
\midrule
weights\_AIC & Model weights derived from AIC.\\
model\_best & Model selected using the minimum AIC.\\
mu\_best, pr\_best, haz\_best & Selected-model estimates.\\
mu\_gen, pr\_gen, haz\_gen & Generating-model estimates, when a reference model is specified.\\
mu\_aic, pr\_aic, haz\_aic & AIC-weighted model-averaged estimates.\\
\bottomrule
\end{tabular}
\end{center}

\hypertarget{bootstrap-confidence-intervals}{%
\subsection{Bootstrap confidence
intervals}\label{bootstrap-confidence-intervals}}

Bootstrap confidence intervals are obtained using:

\texttt{confint(object,\ data,\ B,\ BB,\ level)}

In addition to the fitted object returned by \texttt{marp()},
\texttt{confint()} uses the arguments shown in Table S6.

\textbf{Table S6: Additional arguments used by \texttt{confint()}.}

\begin{center}
\small

\begin{tabular}{ll}
\toprule
Argument & Description\\
\midrule
data & Observed inter-event times; required because fitted objects do not store the data.\\
B & Number of bootstrap resamples.\\
BB & Number of double-bootstrap resamples used for Studentised intervals.\\
level & Confidence level used to construct the confidence intervals.\\
\bottomrule
\end{tabular}
\end{center}

The function returns bootstrap-based model-selection frequencies,
bootstrap-weighted estimates and confidence interval limits for the mean
inter-event time, event probability and hazard function. The main
returned objects are summarised in Table S7.

\begin{samepage}
\textbf{Table S7: Main objects returned by \texttt{confint()}.}

\begin{center}
\footnotesize

\begin{tabular}{p{0.21\linewidth}p{0.43\linewidth}p{0.30\linewidth}}
\toprule
Output type & Returned objects & Description\\
\midrule
Bootstrap model weights & weights\_bstrp & Bootstrap model-selection frequencies.\\
Bootstrap-weighted estimates & mu\_bstrp, pr\_bstrp, haz\_bstrp & Bootstrap-weighted estimates of the mean, event probability and hazard.\\
Percentile CI: generating model & mu\_gen, mu\_gen\_lower, mu\_gen\_upper;\newline pr\_gen, pr\_gen\_lower, pr\_gen\_upper;\newline haz\_gen, haz\_gen\_lower, haz\_gen\_upper & Median and percentile limits for the generating-model estimates.\\
Percentile CI: selected model & mu\_best, mu\_best\_lower, mu\_best\_upper;\newline pr\_best, pr\_best\_lower, pr\_best\_upper;\newline haz\_best, haz\_best\_lower, haz\_best\_upper & Median and percentile limits for the selected-model estimates.\\
Studentised CI: generating model & mu\_lower\_gen, mu\_upper\_gen;\newline pr\_lower\_gen, pr\_upper\_gen;\newline haz\_lower\_gen, haz\_upper\_gen & Studentised limits for generating-model estimates.\\
\addlinespace
Studentised CI: selected model & mu\_lower\_best, mu\_upper\_best;\newline pr\_lower\_best, pr\_upper\_best;\newline haz\_lower\_best, haz\_upper\_best & Studentised limits for selected-model estimates.\\
Studentised CI: model averaged & mu\_lower\_ma, mu\_upper\_ma;\newline pr\_lower\_ma, pr\_upper\_ma;\newline haz\_lower\_ma, haz\_upper\_ma & Studentised limits for model-averaged estimates.\\
\bottomrule
\end{tabular}
\end{center}
\end{samepage}

\hypertarget{simulation-results}{%
\section{Simulation results}\label{simulation-results}}

This supplementary material provides condensed numerical summaries
supporting the simulation results in Section 4.

\hypertarget{point-estimate}{%
\subsection{Point estimate}\label{point-estimate}}

Table S8 reports point-estimation performance for the mean inter-event
time. The table focuses on \(\log_{10}(\mathrm{RMSE})\), complementing
Figure 1 in the main text.

\textbf{Table S8: Point-estimation performance for the mean inter-event
time. The table reports \(\log_{10}(\mathrm{RMSE})\) across simulation
scenarios, sample sizes and estimation methods. Generating-model
estimates are reported only for the well-specified Gamma and Weibull
scenarios.}

\begin{center}
\small

\begin{tabular}{lllll}
\toprule
Scenario & Method & n = 20 & n = 50 & n = 100\\
\midrule
Gamma & Model averaging–AIC & 1.598 & 1.403 & 1.246\\
Gamma & Best model–AIC & 1.596 & 1.399 & 1.241\\
Gamma & Generating model & 1.593 & 1.394 & 1.238\\
Weibull & Model averaging–AIC & 2.943 & 3.306 & 2.604\\
Weibull & Best model–AIC & 2.537 & 2.304 & 2.004\\
\addlinespace
Weibull & Generating model & 2.098 & 1.897 & 1.719\\
Mixture & Model averaging–AIC & 2.376 & 1.987 & 1.925\\
Mixture & Best model–AIC & 2.172 & 2.031 & 1.954\\
\bottomrule
\end{tabular}
\end{center}

The largest errors occurred in the challenging Weibull scenario,
consistent with the stronger right-skewness of the data-generating
distribution. In the Gamma scenario, the three estimators gave very
similar performance, while in the Mixture scenario the two fitted-model
approaches were broadly comparable.

\hypertarget{confidence-intervals}{%
\subsection{Confidence intervals}\label{confidence-intervals}}

Table S9 summarises conditional confidence-interval coverage across the
four target quantities considered in the simulation study. Coverage is
calculated among successful confidence-interval computations, so the
reported coverage values are conditional on successful interval
construction. The failure rate summarises the proportion of replications
in which a valid interval was not returned.

Across scenarios, Model averaging--AIC generally had higher or
comparable conditional coverage than Best model--AIC, particularly for
hazard estimands. These summaries should be interpreted together with
the reported failure rates because coverage is calculated only among
successful confidence-interval computations.

\textbf{Table S9: Conditional confidence-interval coverage summary
across the four target quantities: mean inter-event time, event
probability \(\Pr(X \leq y)\), and hazards at \(t = 150\) and
\(t = 200\). The table reports the range and mean of empirical coverage
across the four targets, together with the overall confidence-interval
failure rate for each scenario and sample size.}

\begin{center}
\small

\begin{tabular}{llllll}
\toprule
Scenario & Method & n & Coverage range & Mean coverage & Failure rate\\
\midrule
Gamma & Model averaging–AIC & 50 & 0.954–0.966 & 0.959 & 0.000\\
Gamma & Best model–AIC & 50 & 0.881–0.947 & 0.910 & 0.000\\
Gamma & Generating model & 50 & 0.941–0.953 & 0.947 & 0.000\\
Gamma & Model averaging–AIC & 100 & 0.946–0.957 & 0.950 & 0.000\\
Gamma & Best model–AIC & 100 & 0.849–0.953 & 0.894 & 0.000\\
\addlinespace
Gamma & Generating model & 100 & 0.943–0.950 & 0.947 & 0.000\\
Weibull & Model averaging–AIC & 50 & 0.939–0.969 & 0.952 & 0.093\\
Weibull & Best model–AIC & 50 & 0.906–0.944 & 0.922 & 0.093\\
Weibull & Generating model & 50 & 0.942–0.958 & 0.954 & 0.093\\
Weibull & Model averaging–AIC & 100 & 0.934–0.951 & 0.945 & 0.093\\
\addlinespace
Weibull & Best model–AIC & 100 & 0.901–0.937 & 0.918 & 0.093\\
Weibull & Generating model & 100 & 0.937–0.951 & 0.946 & 0.093\\
Mixture & Model averaging–AIC & 50 & 0.867–0.967 & 0.930 & 0.015\\
Mixture & Best model–AIC & 50 & 0.798–0.952 & 0.884 & 0.015\\
Mixture & Model averaging–AIC & 100 & 0.827–0.972 & 0.906 & 0.024\\
\addlinespace
Mixture & Best model–AIC & 100 & 0.782–0.966 & 0.868 & 0.024\\
\bottomrule
\end{tabular}
\end{center}

\hypertarget{application-results}{%
\section{Application results}\label{application-results}}

This supplementary material provides numerical summaries supporting the
two application examples in Section 5. The main text presents the fitted
model-support and model-averaged hazard figures, while the tables below
report the corresponding model-comparison and log-hazard estimates.

\hypertarget{model-comparison}{%
\subsection{Model comparison}\label{model-comparison}}

Table S10 reports AIC values, AIC differences and AIC weights for the
six candidate waiting-time models fitted to each application dataset.

\begin{samepage}
\textbf{Table S10: Model comparison for the two application datasets.
Six candidate waiting-time models were fitted using \texttt{marp}. The
table reports the Akaike information criterion (AIC), the AIC difference
from the minimum-AIC model, \(\Delta \mathrm{AIC}\), and the
corresponding AIC weight.}

\begin{center}
\small

\begin{tabular}{lllll}
\toprule
Dataset & Model & AIC & $\Delta$AIC & AIC weight\\
\midrule
Seizure & BPT & 271.11 & 0.00 & 0.315\\
Seizure & Log-normal & 271.53 & 0.43 & 0.255\\
Seizure & Gamma & 272.24 & 1.13 & 0.179\\
Seizure & Log-logistic & 272.89 & 1.79 & 0.129\\
Seizure & Weibull & 273.10 & 2.00 & 0.116\\
\addlinespace
Seizure & Poisson & 279.07 & 7.96 & 0.006\\
Airplane glass & BPT & 211.90 & 0.00 & 0.301\\
Airplane glass & Log-normal & 212.02 & 0.12 & 0.284\\
Airplane glass & Gamma & 212.23 & 0.33 & 0.256\\
Airplane glass & Log-logistic & 214.23 & 2.33 & 0.094\\
\addlinespace
Airplane glass & Weibull & 214.98 & 3.07 & 0.065\\
Airplane glass & Poisson & 276.53 & 64.62 & 0.000\\
\bottomrule
\end{tabular}
\end{center}
\end{samepage}

For both application datasets, the BPT model had the smallest AIC, but
AIC weights were distributed across several flexible two-parameter
models. This supports reporting model-averaged summaries alongside the
selected-model results.

\hypertarget{log-hazard-estimates-and-empirical-comparisons}{%
\subsection{Log-hazard estimates and empirical
comparisons}\label{log-hazard-estimates-and-empirical-comparisons}}

Table S11 reports Best model--AIC and Model averaging--AIC log-hazard
estimates at selected empirical quantiles of each application dataset.
The lower and upper limits correspond to the Model averaging--AIC
bootstrap confidence intervals obtained using \texttt{confint()} with
\(B = BB = 999\). These summaries show that the selected-model and
model-averaged estimates are similar but not identical, especially at
later time points where model-specific hazard shapes can differ.

\textbf{Table S11: Best model--AIC and Model averaging--AIC log-hazard
estimates for the two application datasets. Estimates are evaluated at
selected empirical quantiles of the observed waiting-time distribution.
The lower and upper limits give the Model averaging--AIC bootstrap
confidence interval obtained using \texttt{confint()} with
\(B = BB = 999\).}

\begin{center}
\small

\begin{tabular}{lllllll}
\toprule
Dataset & Quantile & Time & Best model–AIC & Model averaging–AIC & Lower & Upper\\
\midrule
Seizure & 10\% & 7403.20 & -9.692 & -9.774 & -10.718 & -8.970\\
Seizure & 36.7\% & 9915.80 & -9.324 & -9.404 & -10.164 & -8.851\\
Seizure & 63.3\% & 17221.60 & -9.029 & -9.030 & -9.865 & -8.483\\
Seizure & 90\% & 28318.00 & -8.976 & -8.928 & -9.930 & -8.227\\
Airplane glass & 10\% & 23.03 & -3.102 & -3.151 & -3.733 & -2.653\\
\addlinespace
Airplane glass & 23.3\% & 25.50 & -2.661 & -2.708 & -3.189 & -2.325\\
Airplane glass & 36.7\% & 26.77 & -2.490 & -2.529 & -2.986 & -2.164\\
Airplane glass & 50\% & 29.90 & -2.174 & -2.192 & -2.631 & -1.831\\
Airplane glass & 63.3\% & 33.76 & -1.920 & -1.915 & -2.366 & -1.564\\
Airplane glass & 76.7\% & 35.91 & -1.819 & -1.803 & -2.271 & -1.442\\
\addlinespace
Airplane glass & 90\% & 39.58 & -1.689 & -1.659 & -2.177 & -1.245\\
\bottomrule
\end{tabular}
\end{center}

Supplementary Figures S1 and S2 provide graphical diagnostic comparisons
of the empirical, Best model--AIC and Model averaging--AIC log-hazard
curves. These figures are included to show how the fitted summaries
relate to the empirical log-hazard pattern in each dataset. They should
be interpreted descriptively rather than as formal evidence that one
estimator is uniformly superior, because empirical hazard estimates can
be noisy, particularly in small samples.

\begin{center}\includegraphics[width=1\linewidth,alt={Line plot comparing empirical, best-model AIC, and model-averaged AIC log-hazard estimates for the seizure recurrence data.}]{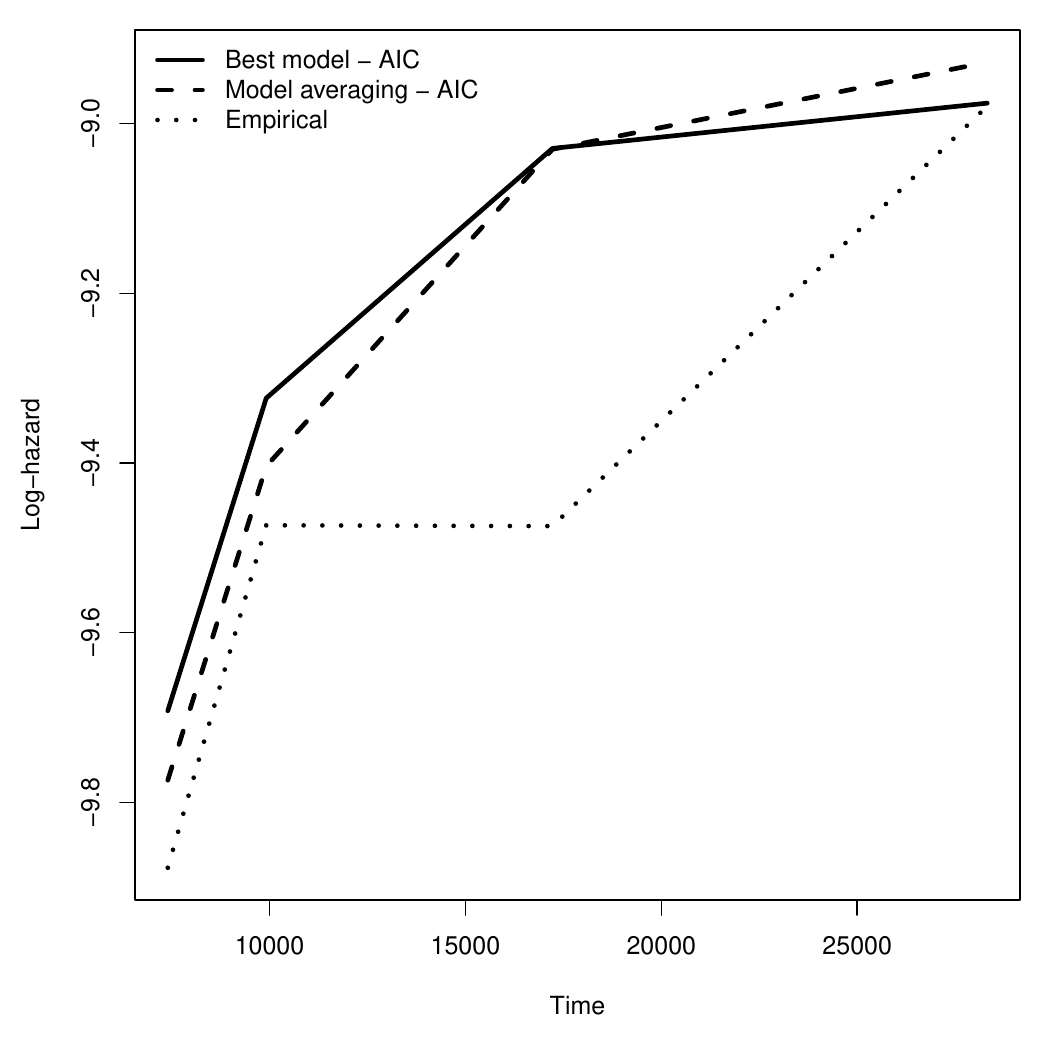} \end{center}

\noindent Figure S1: Empirical, Best model--AIC and Model averaging--AIC
log-hazard estimates for the seizure recurrence data. The empirical
curve is included as a descriptive diagnostic comparison with the fitted
log-hazard summaries. The plot should be interpreted cautiously because
the empirical curve is based on a small number of inter-event times.

\begin{center}\includegraphics[width=1\linewidth,alt={Line plot comparing empirical, best-model AIC, and model-averaged AIC log-hazard estimates for the airplane glass failure-time data.}]{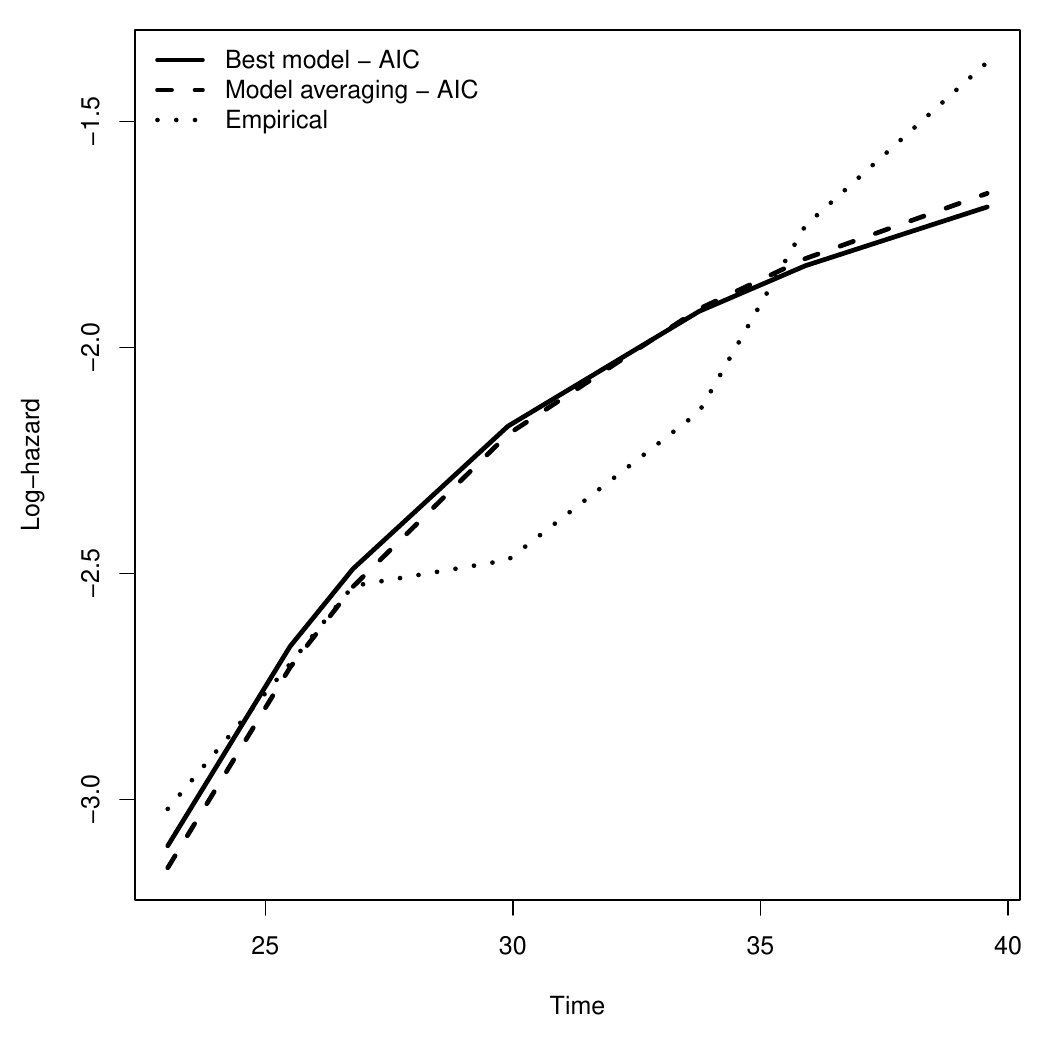} \end{center}

\noindent Figure S2: Empirical, Best model--AIC and Model averaging--AIC
log-hazard estimates for the airplane glass failure-time data. The
empirical curve is included as a descriptive diagnostic comparison with
the fitted log-hazard summaries.

\end{document}